\documentclass[11pt,a4paper]{article}
\pdfoutput=1 
\usepackage{jheppub}
\usepackage{graphicx}
\usepackage{bm}
\usepackage{amsmath}
\usepackage{amsfonts}
\usepackage{braket}
\usepackage{amssymb}
\usepackage{epstopdf}
\usepackage{hyperref}
\usepackage{mathrsfs}
\usepackage{subfigure}
\usepackage{slashed}
\usepackage{color}
\definecolor{green}{rgb}{0.16, 0.5, 0}

\begin{document}
\preprint{MAN/HEP/2014/10, 
CERN-PH-TH/2014-150}

\title{Maximally Symmetric Two Higgs Doublet Model with Natural
  Standard Model Alignment}
\author[a] {P. S. Bhupal Dev}
\author[a,b]{and Apostolos Pilaftsis} 
\affiliation[a]{Consortium for Fundamental Physics, School of Physics and
  Astronomy, \\ 
University of Manchester, Manchester, M13 9PL, United
  Kingdom}  
\affiliation[b]{CERN, Department of Physics, Theory Division, CH-1211
  Geneva 23, Switzerland} 

\abstract{
We study the Higgs mass spectrum as predicted by a Maximally Symmetric
Two Higgs Doublet Model (MS-2HDM) potential based on the ${\rm SO(5)}$
group, softly  broken by bilinear Higgs  mass terms. We  show that the
lightest Higgs  sector resulting  from this MS-2HDM  becomes naturally
aligned with  that of  the Standard Model  (SM), independently  of the
charged   Higgs   boson   mass   and   $\tan   \beta$.  In the context of 
Type-II 2HDM, SO(5) is the simplest of the {\em three} possible symmetry realizations 
of the scalar potential that can naturally lead to the SM alignment. 
Nevertheless,
renormalization group  effects due to  the hypercharge  gauge coupling
$g'$ and third-generation Yukawa  couplings may break sizeably this 
alignment in the MS-2HDM, along  with the custodial  symmetry inherited by  the ${\rm
  SO(5)}$ group.   Using the current  Higgs signal strength  data from
the LHC, which disfavour large deviations from the SM alignment limit,
we derive lower mass bounds on the heavy Higgs sector as a function of
$\tan\beta$, which can be stronger than the existing limits for a wide
range of parameters.  In particular,  we propose a new collider signal
based  on the observation  of four  top quarks  to directly  probe the
heavy Higgs sector of the MS-2HDM during the run-II phase of the LHC.}

\maketitle

\section{Introduction}\label{sec:intro}

The discovery  of a  Higgs resonance with  mass around 125~GeV  at the
LHC~\cite{discovery1, discovery2} offers  an unprecedented opportunity  for probing
extended Higgs scenarios beyond the Standard Model~(SM).  Although the
measured  properties of  the  discovered Higgs  boson show  remarkable
consistency   with  those   predicted   by  the   SM~\cite{couplings1,
  couplings2},  the current  experimental  data still  leave open  the
possibility of new physics that results from an extended Higgs sector.
In  fact,  several  well-motivated  new-physics scenarios  require  an
enlarged Higgs  sector, such as  supersymmetry~\cite{hunter}, in order
to address a number  of theoretical and cosmological issues, including
the gauge  hierarchy problem,  the origin of  the Dark Matter  and the
baryon asymmetry in  our Universe.  Here we follow  a modest bottom-up
approach and  consider one of the simplest  Higgs-sector extensions of
the SM, namely the Two Higgs Doublet Model~(2HDM)~\cite{review}.

The  2HDM   contains  two   complex  scalar  fields   transforming  as
iso-doublets   $({\bf  2},1)$   under   the  SM electroweak  gauge  group   ${\rm
  SU(2)}_L\otimes  {\rm  U(1)}_Y$:  
\begin{eqnarray}
\Phi_i \ = \  \left(\begin{array}{c}
  \phi_i^+ \\  \phi_i^0 \end{array}\right)\; , 
\end{eqnarray}
 with $i=1,2$.  In
this  doublet field  space  $\Phi_{1,2}$, the  general 2HDM  potential
reads
\begin{align}
  \label{pot}
V \ = \ & -\mu_1^2(\Phi_1^\dag \Phi_1) -\mu_2^2 (\Phi_2^\dag \Phi_2) 
-\left[m_{12}^2 (\Phi_1^\dag \Phi_2)+{\rm H.c.}\right]\nonumber\\
& +\lambda_1(\Phi_1^\dag \Phi_1)^2+\lambda_2(\Phi_2^\dag \Phi_2)^2
 +\lambda_3(\Phi_1^\dag \Phi_1)(\Phi_2^\dag \Phi_2) 
 +\lambda_4(\Phi_1^\dag \Phi_2)(\Phi_2^\dag \Phi_1)\nonumber \\
& 
+\left[\frac{1}{2}\lambda_5(\Phi_1^\dag \Phi_2)^2
+\lambda_6(\Phi_1^\dag \Phi_1)(\Phi_1^\dag \Phi_2) 
+\lambda_7(\Phi_1^\dag \Phi_2)(\Phi_2^\dag \Phi_2)+{\rm H.c.}\right]
\; ,
\end{align}
which contains  four real mass  parameters $\mu_{1,2}^2$, Re$(m^2_{12})$,
Im$(m^2_{12})$,  and ten  real  quartic couplings  $\lambda_{1,2,3,4}$,
Re($\lambda_{5,6,7})$,  and Im($\lambda_{5,6,7}$).  As  a consequence,
the   vacuum   structure   of   the   general  2HDM   can   be   quite
rich~\cite{Battye:2011jj},  and in  principle,  can allow  for a  wide
range  of  parameter space  still  compatible  with  the existing  LHC
constraints.    However,   additional   requirements,  such   as   the
Glashow--Weinberg condition~\cite{gw1, gw2}, must be imposed, so as to avoid
Higgs  interactions with  unacceptably large  flavour  changing neutral
currents (FCNC) at the tree level.  The Glashow--Weinberg condition is
satisfied  by four  discrete  choices of  tree-level Yukawa  couplings
between the Higgs doublets and  SM fermions.\footnote{In general, the absence of tree-level flavour-changing couplings of the neutral scalar fields can be guaranteed by requiring the Yukawa coupling matrices to be aligned in flavour space~\cite{pich}.} By performing global fits
to the current  Higgs signals at the LHC and Tevatron  in terms of the
2HDM   parameter   space,   it   has  been   shown~\cite{fit1,   fit2, fit3, fit4, fit5, fit6, 
fit7, 
  Craig:2013hca} that all four  discrete 2HDM types are constrained to
lie close to the so-called SM {\it alignment limit}, in which the mass
eigenbasis of the  CP-even scalar sector aligns with  the SM gauge
eigenbasis.   Specifically,  in  the  Type-II  (MSSM-type)  2HDM,  the
coupling of the  SM-like Higgs to vector bosons  is constrained to lie
within  10\%   of  the   SM  value  at   95\%  CL~\cite{Craig:2013hca, Grinstein:2013npa, 
  Eberhardt:2013uba, eber2, Chiang:2013ixa}.

In light  of the  present and upcoming  LHC data,  possible mechanisms
that lead  to the SM alignment  limit within the  2HDM require further
investigation and scrutiny.  Naively,  the SM alignment limit is often
associated with  the decoupling limit,  in which all  the non-standard
Higgs bosons are assumed to be much heavier than the electroweak scale
so  that the  lightest CP-even  scalar behaves  like the  SM Higgs
boson.   This  SM  alignment  limit  can  also  be  achieved,  without
decoupling~\cite{Chankowski:2000an, Gunion:2002zf,    Ginzburg:2004vp,   Carena:2013ooa}.\footnote{A similar situation was also discussed in an 
extension of the MSSM with a triplet scalar field~\cite{Delgado:2013zfa}, where 
alignment without decoupling could be achieved in a parameter region at small $\tan\beta\lesssim 10$.}
However, for  small $\tan\beta$ values, this is  usually attributed to
accidental cancellations in the 2HDM potential~\cite{Carena:2013ooa}.

In this paper,  we seek a symmetry of the  2HDM potential to naturally
justify the alignment limit,  without decoupling, independently of the
kinematic parameters  of the  theory, such as  the charged  Higgs mass
and~$\tan\beta$.  We  show that  a Maximally Symmetric  2HDM (MS-2HDM)
potential based on  the ${\rm SO}(5)$ group can  {\it naturally} realize the
alignment limit, where ${\rm SO(5)}$ acts on a bilinear field space to
be discussed in Section~\ref{sec:MS2HDM}. In Section~\ref{sec:RGE}, we show that, 
in the context of Type-II 2HDM, the maximal symmetry group SO(5) is the simplest of the {\em three} 
possible symmetry realizations of the scalar potential having natural alignment.   
Nevertheless, as we analyze
in Section~\ref{sec:RGE}, renormalization group (RG) effects
due to the hypercharge gauge coupling $g'$ and third-generation Yukawa
couplings,   as  well  as   soft-breaking  mass   parameters,  violate
explicitly  the  ${\rm  SO}(5)$  symmetry, thereby  inducing  relevant
deviations   from   the   alignment   limit.    As   we   discuss   in
Section~\ref{sec:MP}, such deviations lead to distinct predictions for
the Higgs  spectrum of the MS-2HDM.   In Section~\ref{sec:signals}, we
present a novel collider signature of the MS-2HDM with four top quarks
as   final  states.    In  Section~\ref{sec:concl}   we   present  our
conclusions.  Finally, several technical  details related to our study
have been relegated to Appendices~\ref{app:spec} and~\ref{app:RGE}.

\section{Maximally Symmetric Two Higgs Doublet Model
  Potential\label{sec:MS2HDM}} 

In order to identify all  accidental symmetries of the 2HDM potential,
it   is   convenient    to   introduce   the   8-dimensional   complex
multiplet~\cite{Battye:2011jj,Nishi:2011gc,Pilaftsis:2011ed}: 
\begin{equation}
  \label{Phi}
{\bf \Phi}\ \equiv\ \left(\begin{array}{c}  
\Phi_1  \\
\Phi_2  \\  
\widetilde{\Phi}_1  \\
\widetilde{\Phi}_2 \end{array}\right)\; ,
\end{equation}
where $\widetilde{\Phi}_{i} =  {\rm i}\sigma^2 \Phi^*_{i}$ (with $i=1,2$) and $\sigma^{2}$  is the second  Pauli matrix. We
should remark  that the complex  multiplet ${\bf \Phi}$  satisfies the
Majorana property~\cite{Battye:2011jj}: ${\bf \Phi} = C {\bf \Phi}^*$,
where  $C  =  \sigma^2  \otimes  \sigma^0  \otimes  \sigma^2$  is  the
charge-conjugation matrix, with $\sigma^0 =
{\bf 1}_{2\times  2}$ being the  identity matrix. In terms of the ${\bf \Phi}$-multiplet, the
following   {\em   null}   6-dimensional   Lorentz   vector   can   be
defined~\cite{Battye:2011jj, Pilaftsis:2011ed}:
\begin{equation}
  \label{RA}
R^A\ \equiv\ {\bf \Phi}^\dag \Sigma^A {\bf \Phi}\; ,
\end{equation}  
where  $A=0,1,...,5$  and  the  six $8\times  8$-dimensional  matrices
$\Sigma^A$ may be expressed in terms of the three Pauli matrices $\sigma^{1,2,3}$, as follows:
\begin{eqnarray}
&& \Sigma^{0,1,3} \ = \ \frac{1}{2}\sigma^0 \otimes \sigma^{0,1,3} \otimes
  \sigma^0, \quad   
\Sigma^2 \ = \ \frac{1}{2}\sigma^3 \otimes \sigma^2 \otimes \sigma^0, \nonumber\\
&& \Sigma^4 \ = \ -\frac{1}{2}\sigma^2 \otimes \sigma^2 \otimes \sigma^0, \quad
\Sigma^5 \ = \ -\frac{1}{2}\sigma^1 \otimes \sigma^2 \otimes \sigma^0. 
\end{eqnarray}
We must  emphasize here
that the bilinear  field space spanned by the  6-vector $R^A$ realizes
an {\em orthochronous} ${\rm SO}(1,5)$ symmetry group.

In terms of  the 6-vector $R^A$ defined in~(\ref{RA}), the  2HDM potential $V$
given in~(\ref{pot}) takes on a simple quadratic form:
\begin{equation}
  \label{potR}
V\ =\ -\, \frac{1}{2}\, M_A\,R^A\: + \: \frac{1}{4}\, L_{AB}\, R^A R^B\; ,
\end{equation}
where $M_A$  and $L_{AB}$ are ${\rm SO}(1,5)$  constant `tensors' that
depend  on the  mass parameters  and quartic  couplings of  the scalar
potential~$V$    and    their   explicit    forms    may   be    found
in~\cite{Pilaftsis:2011ed,Maniatis:2007vn,Ivanov:2007de,Nishi:2007dv}.
Requiring   that    the   SU(2)$_L$   gauge-kinetic    term   of   the
multiplet~{\boldmath $\Phi$}  remains canonical restricts  the allowed
set of  rotations from SO(1,5)  to SO(5),\footnote{We note  in passing
  that if the restriction of SU(2)$_L$ gauge invariance is lifted, the
  2HDM  is then equivalent  to an  {\em ungauged}  theory with  8 real
  scalars and so  the maximal symmetry group becomes  the larger group
  O(8)~\cite{Deshpande:1977rw}.}  where only  the  spatial components
$R^I$ (with $I=1,...,5$) transform, whereas the zeroth component $R^0$
remains invariant.   Consequently, in  the absence of  the hypercharge
gauge coupling $g'$ and fermion Yukawa couplings, the maximal symmetry
group of the 2HDM is $G^R_{\rm  2HDM} = {\rm SO(5)}$.  Given the group
isomorphy  ${\rm SO(5)}  \sim  {\rm Sp}(4)/{\rm Z}_2$, the maximal symmetry group  
of the 2HDM
in      the      original      {\boldmath     $\Phi$}-field      space
is~\cite{Pilaftsis:2011ed}\footnote{In~\cite{Pilaftsis:2011ed},   the  symplectic  group
  ${\rm Sp}(4)$ is denoted as ${\rm SU_M}(4)$, where the 10 generators
  of  the restricted  U(4)  group satisfying  a Majorana  (symplectic)
  condition were  presented.}
\begin{equation}
{\rm G}^{\bf \Phi}_{\rm 2HDM}\ =\ \left({\rm Sp}(4)/{\rm  Z}_2\right) \otimes
{\rm SU(2)}_L\; ,
\label{Gphi}
\end{equation}
in the custodial symmetry limit of vanishing  $g'$ and fermion Yukawa couplings. 
The 
quotient factor Z$_2$ in (\ref{Gphi}) is needed to avoid double covering the group ${\rm G}^{\bf \Phi}_{\rm 2HDM}$ in the {\boldmath $\Phi$}-space. One may note here that the 10 Lie generators of Sp(4) 
may be represented in  the {\boldmath $\Phi$}-space as $K^a = \kappa^a \otimes \sigma^0$ 
(with $a = 0,1,2,\dots,
9$), where
\begin{align}
  \label{eq:Ka}
\kappa^0 \ & = \ \frac{1}{2}\, \sigma^3\otimes \sigma^0\;, \qquad 
\kappa^1 \  = \ \frac{1}{2}\, \sigma^3 \otimes \sigma^1\; ,\nonumber\\
\kappa^2 \ & = \ \frac{1}{2}\, \sigma^0 \otimes \sigma^2\; ,\qquad
\kappa^3 \  =\ \frac{1}{2}\, \sigma^3\otimes \sigma^3\;, \nonumber\\
\kappa^4 \ & = \ \frac{1}{2}\, \sigma^1 \otimes \sigma^0 \; ,\qquad
\kappa^5 \  = \ \frac{1}{2}\, \sigma^1 \otimes \sigma^3 \; ,\\
\kappa^6 \ & = \ \frac{1}{2}\, \sigma^2\otimes \sigma^0 \;, \qquad 
\kappa^7\  = \ \frac{1}{2}\, \sigma^2 \otimes \sigma^3 \; ,\nonumber\\
\kappa^8 \ & = \ \frac{1}{2}\, \sigma^1 \otimes \sigma^1 \; ,\qquad
\kappa^9 \  = \  \frac{1}{2}\, \sigma^2 \otimes \sigma^1 \; ,\nonumber
\end{align}
with   the   normalization:   ${\rm   Tr}  (\kappa^a\,\kappa^b   )   =
\delta^{ab}$. Thus, the group ${\rm G}^{\bf \Phi}_{\rm 2HDM}$ includes the 
U(1)$_Y$ hypercharge group through the Sp(4) generator $K^0$, whereas the 9 other 
Sp(4) generators listed in (\ref{eq:Ka}) are related to various Higgs Family and CP 
transformations~~\cite{Pilaftsis:2011ed}. On the other hand, the SU(2)$_L$ generators in the  
{\boldmath $\Phi$}-space may be written as $\sigma^0\otimes \sigma^0\otimes (\sigma^b/2)$ (with $b=1,2,3$), which manifestly commute with all Sp(4) generators $K^a$.

As we  will see below by an explicit construction~[cf.~(\ref{VSO5}) and Section~\ref{sec:cust}], it is not difficult to deduce that, in the custodial symmetry limit, the maximal symmetry
group for an $n$ Higgs Doublet Model ($n$HDM) will be
\begin{equation}
  \label{GnHDM}
{\rm G}^{\bf \Phi}_{n{\rm HDM}}\ =\ \left({\rm Sp}(2n)/{\rm  Z}_2\right) \otimes
{\rm SU(2)}_L\; ,
\end{equation}
in  which   case  the  multiplet  ${\bf  \Phi}$   becomes  a  Majorana
$4n$-dimensional complex  vector.\footnote{Given an apparently deep connection between SO($2n+1$) and Sp($2n$) groups~\cite{Patera:1979dk, Slansky:1981yr}, both of which have $n(2n+1)$ generators, one might be able to identify the necessary bilinears in the $R$-space for any $n$HDM. However, this is somewhat non-trivial for $n\geq 3$, and therefore, we postpone this discussion 
to a future dedicated study.}
It is  interesting to note  that for
the SM  with $n=1$ Higgs doublet, (\ref{GnHDM})  yields the well-known
result:  ${\rm  G}^{\bf \Phi}_{\rm  SM}  =  ({\rm SU}(2)_C/{\rm  Z}_2)
\otimes {\rm SU(2)}_L$, by virtue of the group isomorphy: ${\rm Sp}(2)
\sim {\rm  SU}(2)_C$, where ${\rm SU}(2)_C$ is  the custodial symmetry
group  originally introduced  in~\cite{Sikivie:1980hm}.  Hence,  it is
important  to stress  that (\ref{GnHDM})  represents a  general result
that holds for any $n$HDM.

We may  now identify all maximal  symmetries of the  2HDM potential by
classifying all proper, improper and semi-simple subgroups of SO(5) in
the    bilinear    $R^I$    space.     In    this    way,    it    was
found~\cite{Battye:2011jj,Pilaftsis:2011ed}  that   a  2HDM  potential
invariant  under ${\rm  SU(2)}_L\otimes  {\rm U(1)}_Y$  can possess  a
maximum  of 13  accidental symmetries.   This  symmetry classification
extends    the   previous    list   of    six    symmetries   reported
in~\cite{Ivanov:2007de},  where possible  custodial symmetries  of the
theory were not  included.  Each of the 13  classified symmetries puts
some restrictions  on the kinematic  parameters appearing in  the 2HDM
potential~(\ref{pot}).   In  a  specific diagonally  reduced  bilinear
basis~\cite{Gunion:2005ja,Maniatis:2011qu},   one   has  the   general
restrictions   Im$(\lambda_5)=0$   and   $\lambda_6=\lambda_7$,   thus
reducing  the number of  independent quartic  couplings to seven.   In the
maximally  symmetric SO(5)  ($\sim {\rm  Sp}(4)/{\rm Z}_2$)  limit, we
have   the   following   relations   between  the   scalar   potential
parameters~\cite{Battye:2011jj,Pilaftsis:2011ed}:
\begin{align}
& \mu_1^2 \ = \ \mu_2^2\; , \quad m^2_{12} \ = \ 0\; , \quad \nonumber \\
& \lambda_2 \ = \ \lambda_1\; , \quad 
 \lambda_3  \ = \ 2\lambda_1\; , \quad 
\lambda_4 \ = \  {\rm Re}(\lambda_5) \  = \  \lambda_6 \ = \ \lambda_7\ =\ 0 \; .
\label{so5}
\end{align} 
Thus,  in   the  SO(5)  limit,  the  2HDM   potential  (\ref{pot})  is
parametrized  by a  {\em single}  mass  parameter $\mu^2$  and a  {\em
  single} quartic coupling $\lambda$:
\begin{align}
   \label{VSO5}
V \ & = \  -\,\mu^2\, \Big(|\Phi_1|^2+|\Phi_2|^2\Big)\: +\: \lambda\,
\Big(|\Phi_1|^2+|\Phi_2|^2\Big)^2\nonumber\\
 \ & = \ -\: \frac{\mu^2}{2}\, {\bf \Phi}^\dagger\, {\bf \Phi}\ +\ 
\frac{\lambda}{4}\, \big( {\bf \Phi}^\dagger\, {\bf \Phi}\big)^2   \; .
\end{align}
It   is   worth   stressing   that  the   MS-2HDM   scalar   potential
in~(\ref{VSO5}) is  more minimal than the respective  potential of the
MSSM at the  tree level. Even in the  custodial symmetric limit $g'\to
0$, the latter only  possesses a smaller symmetry: ${\rm O}(2)\otimes
{\rm  O}(3)  \subset  {\rm  SO}(5)$,  in  the  5-dimensional  bilinear
$R^I$~space.

\subsection{Custodial Symmetries in the MS-2HDM}\label{sec:cust}

It  is  now  interesting  to  discuss the  implications  of  custodial
symmetries for the Yukawa sector of the 2HDM. To this end, let us only
consider  the quark Yukawa  sector of  the theory,  even though  it is
straightforward to  extend our results  to the lepton sector  as well.
The  relevant part  of the  quark-Yukawa  Lagrangian in  the 2HDM  can
generally be written down as follows:
\begin{align}
-{\cal L}^q_Y \ & = \  \bar{Q}_L(h_1^u \widetilde{\Phi}_1+ h_2^u\widetilde{\Phi}_2)u_R \: +
\: \bar{Q}_L(h_1^d {\Phi}_1+ h_2^d {\Phi}_2)d_R \;
\nonumber\\ 
\ & =\; \left(\bar{u}_L\,,\, \bar{d}_L\right)\; \left(\widetilde{\Phi}_1 \; , \ \widetilde{\Phi}_2 \; , \Phi_1 \; ,
\ \Phi_2 \right)\,
{\cal H}\,  \left(\begin{array}{c} u_R \\ d_R \end{array}
\right) \; ,\qquad \nonumber
\end{align}
where $Q_L \equiv (u_L\, ,\, d_L)^{\sf T}$ is the SM quark iso-doublet and 
we  have  introduced  a  $12\times  6$-dimensional  {\em
  non}-square  Yukawa  coupling matrix
\begin{equation}
  \label{calH}
{\cal H}\ \equiv\ 
\left(\begin{array}{ll} h_1^u & {\bf 0}_{3\times 3} \\
h_2^u & {\bf 0}_{3\times 3} \\
{\bf 0}_{3\times 3} & h_1^d \\
{\bf 0}_{3\times 3} & h_2^d 
\end{array}\right)\; .
\end{equation}

All the custodial symmetries of the 2HDM potential can be deduced by examining  
the Sp(4)  generators $K^a=\kappa^a\otimes \sigma^0$ in  the {\boldmath $\Phi$}-space, where 
$\kappa^a$ are explicitly given in~(\ref{eq:Ka}). Candidate
Sp(4) generators  of the custodial symmetry are  those generators that
do {\em not} commute  with the hypercharge generator $K^0$, i.e.~$K^a$
with  $a =  4,5,6,7,8,9$.  It  is not  difficult to  see that  these six
generators, together with $K^0$, form {\it three} inequivalent realizations of
the      SU(2)$_C$     custodial     symmetry~\cite{Pilaftsis:2011ed}:
(i)~$K^{0,4,6}$, (ii)~$K^{0,5,7}$ and (iii)~$K^{0,8,9}$.

In   order  to   see   the  implications   of   the  three   custodial
symmetries~(i),~(ii) and~(iii) for the  quark Yukawa sector, we impose
a symmetry  commutation relation on  ${\cal H}$ after  generalizing it
for {\em non}-square matrices:
\begin{equation}
  \label{CCond}
\kappa^a\, {\cal H}\: -\:  {\cal H}\, t^b\ =\ {\bf  0}_{4\times 2}\; ,
\end{equation}
where ${\cal  H}$ is expressed in the  reduced $4\times 2$-dimensional
space, in which the $3\times  3$ flavour space has been suppressed. In
addition, we denote with $t^b =  \sigma^b/2$ (with $b = 1,2,3$) 
the three $2\times 2$ generators of the custodial SU(2)$_C$ group.  
One can immediately check  that it holds $\kappa^0\,  {\cal H} -
{\cal H}\, t^3 = {\bf 0}_{4\times 2}$, which implies that the specific
block  structure of  ${\cal H}$  in~(\ref{calH}) respects  U(1)$_Y$ by
construction,  given  the  correspondence:  $\kappa^0  \leftrightarrow
t^3$.   In  detail,  imposing~(\ref{CCond})  for the  three  SU(2)$_C$
symmetries, we  obtain the following  relations among the  $3\times 3$
up- and down-type quark Yukawa coupling matrices:
\begin{eqnarray}
  \label{Csol}
\mbox{(i)}   &&\quad  
           h^u_1\ =\ e^{i\theta} h^d_1 \quad \mbox{and}\quad 
h^u_2\ =\ e^{i\theta} h^d_2\;, \nonumber\\
\mbox{(ii)}  &&\quad h^u_1\ =\ e^{i\theta} h^d_1 \quad \mbox{and}\quad
h^u_2\ =\ - e^{i\theta} h^d_2\; , \\
\mbox{(iii)} &&\quad h^u_1\ =\ e^{i\theta} h^d_2 \quad \mbox{and}\quad 
h^u_2\ =\ e^{-i\theta} h^d_1\; ,   \nonumber
\end{eqnarray}
where  $\theta$ is  an  arbitrary angle  unspecified  by the  symmetry
constraint~(\ref{CCond}).   We should  stress  again that  only for  a
fully   SO(5)-symmetric   2HDM,    the   three   sets   of   solutions
in~(\ref{Csol}) are equivalent.  However,  this is not in general true
for  scenarios  that  happen  to  realize  only  subgroups  of  SO(5),
according      to      the      symmetry     classification      given
in~\cite{Battye:2011jj,Pilaftsis:2011ed}.

\subsection{Scalar Spectrum in the MS-2HDM}\label{sec:spec-ms}

The masses and mixing in the  Higgs sector of a general 2HDM are given
in  Appendix~\ref{app:spec}.  After  electroweak symmetry  breaking in
the MS-2HDM, we have the breaking pattern
\begin{equation}
{\rm  SO}(5) \  \xrightarrow{\langle \Phi_{1,2}\rangle \neq 0}  \ {\rm
  SO}(4)\; , 
\end{equation} 
which  gives rise  to a  Higgs boson  $H$ with  mass $M_H^2=2\lambda_2
v^2$, whilst  the remaining four  scalar fields, denoted  hereafter as
$h$,  $a$ and  $h^\pm$, are  massless (pseudo)-Goldstone  bosons.  The
latter is  a consequence of the  Goldstone theorem~\cite{goldstone} 
and  can be readily
verified by  means of~(\ref{so5}) in (\ref{mass}).   Thus, we identify
$H$ as  the SM-like Higgs  boson with the mixing  angle $\alpha=\beta$
[cf.~(\ref{HSM})].  We  call this the SM {\it  alignment limit}, which
can be naturally attributed to the SO(5) symmetry of the theory.  

In the exact SO(5)-symmetric limit, the scalar spectrum of the MS-2HDM
is experimentally unacceptable,  as the four massless pseudo-Goldstone
particles, viz.~$h$,  $a$ and $h^\pm$,  have sizeable couplings to  the 
SM $Z$ and $W^\pm$ bosons [cf.~(\ref{coup2})].  
These couplings induce additional decay channels,
such as~$Z\to  ha$ and $W^\pm  \to h^\pm h$, which  are experimentally
excluded~\cite{PDG}. Nevertheless, as we will see in the next section,
the   SO(5)  symmetry  of   the  original   theory  may   be  violated
predominantly by  RG effects due to $g'$  and third-generation Yukawa
couplings, as well as by soft SO(5)-breaking mass parameters, thereby 
lifting the masses of these pseudo-Goldstone particles.

\section{RG and Soft Breaking Effects \label{sec:RGE}}

\begin{figure*}[t]
\centering
\includegraphics[width=7cm]{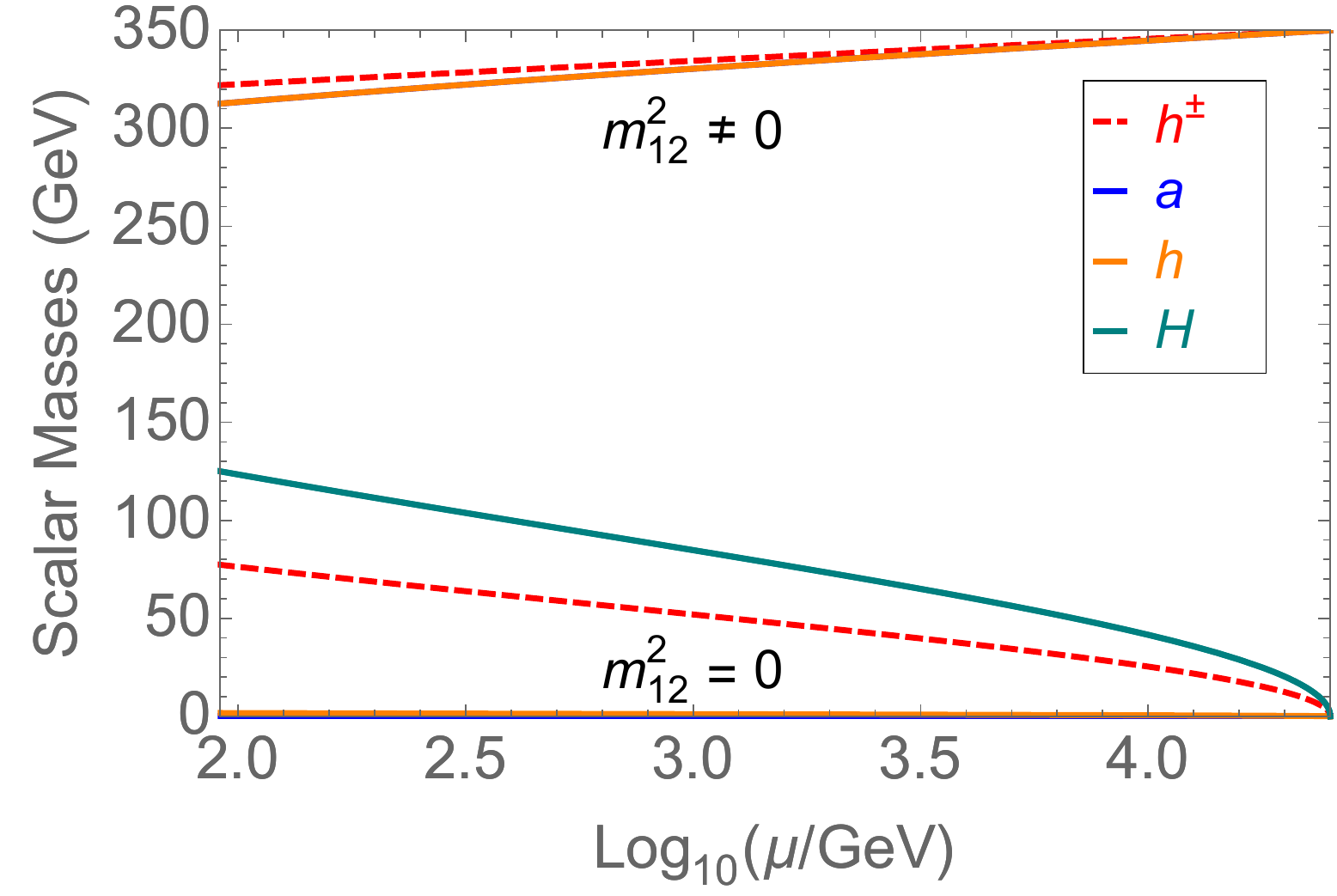}
\hspace{0.5cm}
\includegraphics[width=7cm]{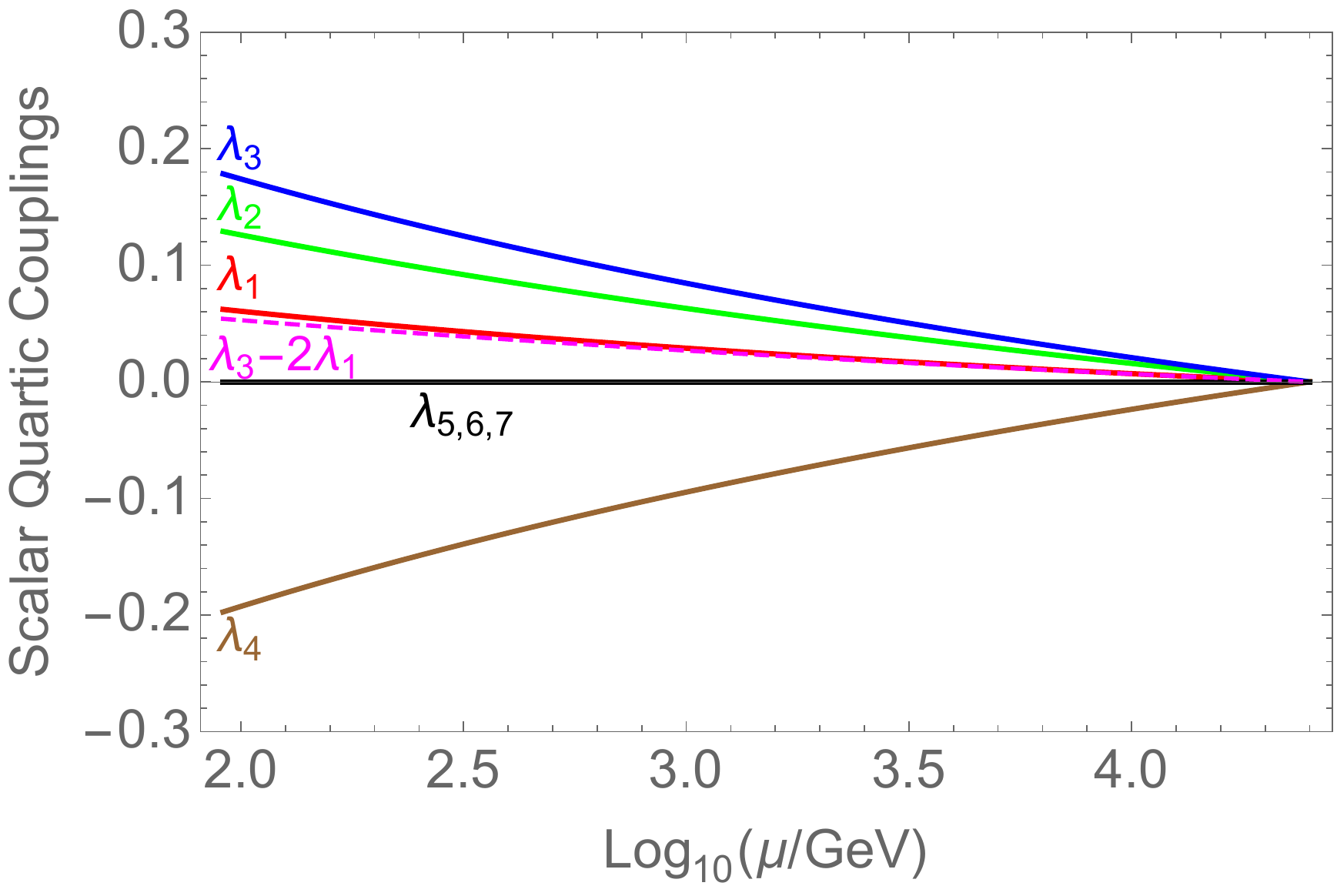}
\caption{(left panel)  The Higgs spectrum  in the MS-2HDM  without and
  with soft breaking effects  induced by $m^2_{12}$. For $m^2_{12}=0$, the pseudo-Goldstone boson $a$ remains massless at tree-level, whereas $h$ and $h^\pm$ receive small masses due to the $g'$ and Yukawa coupling effects. For $m^2_{12}\neq 0$, one obtains a quasi-degenerate heavy Higgs spectrum, cf.~(\ref{mass-so5}). (right panel)
  The  RG evolution  of the  scalar quartic  couplings under  $g'$ and
  Yukawa coupling  effects. Here we have  chosen $\mu_X=2.5\times 10^{4}$ GeV,
  $\lambda(\mu_X)=0$   and   $\tan\beta   =  50$   for   illustration.
} \label{fig1}
\end{figure*}

As discussed in the previous  section, the SO(5) symmetry that governs
the MS-2HDM  will be broken due  to $g'$ and  Yukawa coupling effects,
similar to the  breaking of custodial symmetry in  the SM.  Therefore,
an interesting question  will be to explore whether  these effects are
sufficient to  yield a  viable Higgs spectrum  at the weak  scale.  To
address this question in a  technically natural manner, we assume that
the  SO(5)  symmetry is  realized  at  some  high scale~$\mu_X$.   The
physical mass  spectrum at the  electroweak scale is then  obtained by
the RG evolution  of the 2HDM parameters given  by (\ref{pot}).  Using
state-of-the-art  two-loop RG equations given  in Appendix~\ref{app:RGE}, 
we  examine the
deviation of the Higgs spectrum  from the SO(5)-symmetric limit due to
$g'$   and  Yukawa   coupling   effects.   This   is  illustrated   in
Figure~\ref{fig1}  for a  typical choice  of parameters  in  a Type-II
realization of the  2HDM, even though the conclusions  drawn from this
figure have more general  applicability.  In particular, we obtain the
following  breaking  pattern starting  from  a ${\rm  SU(2)}_L$-gauged
theory:
\begin{eqnarray}
  \label{SO5breaking}
{\rm  SO}(5) \otimes {\rm SU}(2)_L\ &\xrightarrow{~g'\neq 0~}&\  {\rm
  O}(3)\otimes    {\rm    O}(2) \otimes {\rm SU}(2)_L\ \sim\
{\rm O(3)} \otimes {\rm U(1)}_Y \otimes 
{\rm SU}(2)_L \nonumber\\    
&\xrightarrow{{\rm    Yukawa}}&\
 {\rm O}(2)\otimes {\rm U(1)}_Y  \otimes {\rm SU}(2)_L\
\sim\ {\rm U(1)}_{\rm PQ} \otimes {\rm U(1)}_Y \otimes 
{\rm SU}(2)_L\nonumber\\ 
&\xrightarrow{\langle \Phi_{1,2}\rangle \neq 0}&\ {\rm U(1)_{em}}\;  ,
\end{eqnarray} 
where ${\rm  U(1)}_{\rm em}$ is  the electromagnetic group.   In other
words, RG-induced $g'$ effects only lift the charged Higgs-boson mass
$M_{h^\pm}$, while the corresponding Yukawa coupling effects also lift
slightly  the  mass of  the  non-SM CP-even  pseudo-Goldstone
boson~$h$.  However, they still leave the CP-odd scalar $a$ massless
(see left  panel of Figure~\ref{fig1} for $m^2_{12}=0$),  which can be identified  as a
${\rm  U}(1)_{\rm PQ}$  axion~\cite{Peccei:1977hh}.  The  deviation of
the  scalar quartic  couplings from  the SO(5)-symmetric  limit given
in~(\ref{so5}),  thanks  to  $g'$  and  Yukawa  coupling  effects,  is
illustrated in Figure~\ref{fig1} (right panel) for a simple 
choice of the single quartic coupling $\lambda=0$ at the SO(5)-symmetry scale 
$\mu_X$.   

Figure~\ref{fig1} (left panel) also shows that $g'$ and Yukawa coupling effects are
{\em  not} sufficient to  yield a  viable Higgs  spectrum at  the weak
scale, starting  from a  SO(5)-invariant boundary condition  at some  high scale
$\mu_X$.   To minimally circumvent  this problem,  we need  to include
soft SO(5)-breaking  effects, by assuming  a non-zero value  for ${\rm
  Re}(m_{12}^2)$ in the 2HDM potential (\ref{pot}).   
In the  SO(5)-symmetric limit~(\ref{so5})  for the
scalar  quartic couplings,  but  with ${\rm  Re}(m_{12}^2)\neq 0$,  we
obtain the following mass spectrum [cf.~(\ref{mass})]:
\begin{eqnarray}
M_H^2 \ = \ 2\lambda_2 v^2\; , \qquad M_h^2 \ = \ M_a^2 \ = \ M^2_{h^\pm} \ = \
\frac{{\rm Re}(m^2_{12})}{s_\beta c_\beta} \; ,
\label{mass-so5}
\end{eqnarray}
as  well as  an equality  between  the CP-even  and CP-odd  mixing
angles: $\alpha = \beta$, thus predicting an {\it exact} alignment for
the  SM-like  Higgs  boson  $H$, simultaneously with  
an experimentally allowed heavy Higgs spectra (see left  panel of Figure~\ref{fig1} for $m^2_{12}\neq 0$). Note that in the alignment limit, the heavy Higgs sector is exactly degenerate [cf.~(\ref{mass-so5})] at the SO(5) symmetry-breaking scale, and at the low-energy scale, this degeneracy is mildly broken by the RG effects. Thus, we obtain a quasi-degenerate heavy Higgs spectrum in the MS-2HDM, as illustrated in Figure~\ref{fig1} (left panel). We emphasize that this is a unique prediction of this model, valid even in the non-decoupling limit, and can be used to distinguish it from other 2HDM scenarios. 

From (\ref{mass-so5}), we notice that the  alignment  limit
$\alpha=\beta$  is   independent  of  the   charged  Higgs-boson  mass
$M_{h^\pm}$ and  the value of  $\tan\beta$.  This is  achieved without
{\it decoupling}, i.e.~without the need to consider the mass hierarchy
$M_{h^\pm} \gg  v$. Hence,  in this softly  broken SO(5) 2HDM,  we get
{\em       natural}       SM       alignment,      {\em       without}
decoupling.\footnote{Strictly   speaking,  there   will   be  one-loop
  threshold  corrections to the  effective MS-2HDM  potential, sourced
  from  a non-zero  ${\rm Re}(m^2_{12})$,  which might  lead  to small
  misalignments.   A simple estimate  suggests that  these corrections
  are  of order  $\lambda^2/(16 \pi^2)$  and can  therefore  be safely
  neglected to a good approximation.} 
It is instructive to analyze this last point  in more detail.  In
the general CP-conserving 2HDM, the CP-even scalar mass matrix can
be written down as~\cite{Haber:1993an,Pilaftsis:1999qt}
\begin{eqnarray}
  \label{align2}
M^2_{S} \ & = & \ M_a^2\left(\begin{array}{cc}
s_\beta^2 & -s_\beta c_\beta \\ 
-s_\beta c_\beta & c_\beta^2 
\end{array}\right) 
+ v^2\left(\begin{array}{cc}
2\lambda_1 c^2_\beta + \lambda_5 s^2_\beta + 
2 \lambda_6 s_\beta c_\beta & \lambda_{34}s_\beta c_\beta + 
\lambda_6 c^2_\beta + \lambda_7 s^2_\beta\\  
\lambda_{34}s_\beta c_\beta +\lambda_6 c^2_\beta + \lambda_7 s^2_\beta &  
2\lambda_2 s_\beta^2+ \lambda_5 c^2_\beta 
+ 2 \lambda_7 s_\beta c_\beta \end{array}\right) \nonumber\\ 
& \equiv &\ \left(\begin{array}{cc}
c_\beta & -s_\beta \\ 
s_\beta & c_\beta 
\end{array}\right)\; \widehat{M}^2_S\;  
\left(\begin{array}{cc}
c_\beta & s_\beta \\ 
-s_\beta & c_\beta 
\end{array}\right)\; , 
\end{eqnarray}
where $M^2_a$  is given  in~(\ref{mass}), $\lambda_{34} \equiv  \lambda_3 +
\lambda_4$, and 
\begin{equation}
  \label{M2hatS}
\widehat{M}^2_S\  =\  
\left(\begin{array}{cc}
\widehat{A} & \widehat{C} \\
\widehat{C} & \widehat{B}
\end{array}\right)\; ,
\end{equation}
with 
\begin{align}
\widehat{A} &  =  2v^2 \Big[ c_\beta^4 \lambda_1 
+ s_\beta^2 c_\beta^2 \lambda_{345}  
+ s_\beta^4 \lambda_2\:  +\: 2 s_\beta c_\beta \Big( c^2_\beta \lambda_6 +
s^2_\beta \lambda_7\Big)\Big]\; , \label{ahat} \\
\widehat{B} &  =   M_a^2\: +\: \lambda_5 v^2\: +\: 2 v^2 
\Big[ s^2_\beta c^2_\beta
  \Big(\lambda_1+\lambda_2-\lambda_{345}\Big)\:
-\: s_\beta c_\beta \Big(c^2_\beta - s^2_\beta\Big) \Big(\lambda_6 
- \lambda_7\Big) \Big]\; , \label{bhat} \\
\widehat{C} &  =   
v^2 \Big[ s^3_\beta c_\beta \Big( 2\lambda_2-\lambda_{345}\Big) - 
c^3_\beta s_\beta \Big(2\lambda_1- \lambda_{345}\Big) + c^2_\beta
\Big( 1 - 4 s^2_\beta \Big) \lambda_6 
+ s^2_\beta \Big( 4 c^2_\beta - 1\Big) \lambda_7 \Big]. \label{chat}
\end{align}
Here we have used  the  short-hand  notation:  $\lambda_{345} \equiv  \lambda_3  +
\lambda_4    +    \lambda_5$.     Observe    that    $\widehat{M}^2_S$
in~(\ref{align2}) is the respective  $2\times 2$ CP-even mass matrix
written down in the so-called Higgs eigenbasis~\cite{Georgi, Dono, Lavoura, Botella:1994cs}.

Evidently, the SM alignment limit $\alpha \to \beta$ for the CP-even
scalar mixing  angle $\alpha$  is obtained, provided  the off-diagonal
elements  of   $\widehat{M}^2_S$  in~(\ref{M2hatS})  vanish,  i.e.~for
$\widehat{C} =  0$~\cite{Gunion:2002zf}. From (\ref{chat}), this yields the quartic equation
\begin{eqnarray}
\lambda_7 t_\beta^4 -  (2\lambda_2-\lambda_{345})t_\beta^3 + 3(\lambda_6-\lambda_7)t_\beta^2 + 
(2\lambda_1-\lambda_{345})t_\beta - \lambda_6 \ = \ 0 \; .
\label{align-gen}
\end{eqnarray}
In order to satisfy (\ref{align-gen}) for {\it any} value of $\tan\beta$, the coefficients of the polynomial in $\tan\beta$ must identically vanish.\footnote{Notice that (\ref{align-gen}) is satisfied automatically in the SO(5) limit given in (\ref{so5}).} Imposing this restriction, we conclude that all natural alignment solutions must satisfy the following condition:
\begin{eqnarray}
\lambda_1 \ = \ \lambda_2 \ = \ \lambda_{345}/2\;, \qquad \lambda_6 \ = \ \lambda_7 \ = \ 0\; . 
\label{alcond}
\end{eqnarray}
In particular, for $\lambda_6 = \lambda_7 = 0$, (\ref{align-gen}) has a solution 
\begin{eqnarray}
  \label{tanb}
\tan^2\beta\ =\ \frac{2\lambda_1 - \lambda_{345}}{2\lambda_2 -
\lambda_{345}}\ >\ 0 \; ,
\end{eqnarray}
independent of $M_a$. After some algebra, the simple solution (\ref{tanb}) to our general alignment condition (\ref{align-gen}) can be shown to be equivalent to that derived in~\cite{Carena:2013ooa, Carena:2014nza}.

In the alignment limit,  the two  CP-even  Higgs  masses are  given  by the  diagonal
elements of $\widehat{M}^2_S$ in (\ref{M2hatS}):
\begin{eqnarray}
  \label{MH-align}
M_H^2 \ &=& \ 2v^2(\lambda_1 c_\beta^4+\lambda_{345}s_\beta^2 c_\beta^2 +
\lambda_2 s_\beta^4) \  \equiv \ \lambda_{\rm SM}v^2\; ,\\ 
  \label{Mh-align} 
M_h^2 \ &=& \
M_a^2\: +\:  \lambda_5 v^2\: +\: 
2v^2s_\beta^2c_\beta^2(\lambda_1+\lambda_2-\lambda_{345}) \; . 
\end{eqnarray} 

On the other hand, in the limit  $M_a \gg v$, we can use a seesaw-like
approximation in (\ref{M2hatS}) to obtain
\begin{eqnarray}
  \label{MH2seesaw}
M_H^2 \ & \simeq & \ \lambda_{\rm SM}v^2 - \frac{v^4 s^2_\beta
  c^2_\beta}{M_a^2 + \lambda_5 v^2}\,
\Big[s_\beta^2\Big(2\lambda_2 - \lambda_{345}\Big) 
- c_\beta^2\Big(2\lambda_1 - \lambda_{345}\Big)\Big]^2 \; , \\[2mm]
M_h^2 \ & \simeq & \ M_a^2 + \lambda_5 v^2 \ \gg \ v^2\; . \label{lam5}
\end{eqnarray}
In (\ref{MH2seesaw}) and (\ref{lam5}), we have  also included the possibility of decoupling via
a large $\lambda_5$ coupling~\cite{Ginzburg:2004vp}.  For large values
of $\tan\beta$, e.g.~$\tan\beta  \stackrel{>}{{}_\sim} 10$, we readily
see  that (\ref{MH2seesaw}) reduces  to $M_H^2\simeq  2\lambda_2 v^2$,
which   again    leads       to   a    natural      alignment. 

As noted above, in  the SO(5) symmetric  limit of the conformal  part of
the 2HDM  as given by~(\ref{so5}) and (\ref{mass-so5}),  the SM alignment  is achieved
for   any   value   of    $\tan\beta$ [cf.~(\ref{tanb})].    
In  addition to SO(5), one  may now
wonder whether there are other  classified symmetries of the 2HDM that
lead    to   natural    SM   alignment,    {\em    independently}   of
$\tan\beta$ and $M_a$. According to  the classification given in Table~1
of~\cite{Pilaftsis:2011ed},  we observe  that, in the context of Type-II 2HDM,  
there are {\it only} two other
symmetries which lead to such natural SM alignment by satisfying (\ref{alcond}), viz.\footnote{In Type-I 2HDM, there exists an additional possibility of realizing an exact Z$_2$ symmetry~\cite{Deshpande:1977rw} which leads to an exact alignment, i.e. in the context of the so-called inert 2HDM~\cite{Barbieri:2006dq}.}
\begin{eqnarray}
\mbox{(i)}   &&\quad  
           {\rm O(3)}\otimes {\rm O(2)}: \qquad  \lambda_1 \ = \ \lambda_2 \ = \ \lambda_{34}/2, \quad \lambda_5 \ = \ \lambda_6 \ = \ \lambda_7\ = \ 0\;, \label{other-sym1} \\
\mbox{(ii)}  &&\quad {\rm Z}_2\otimes [{\rm O(2)}]^2: \qquad \lambda_1 \ = \ \lambda_2 \ = \ \lambda_{345}/2, \quad \lambda_6 \ = \ \lambda_7\ = \ 0 \; . 
\label{other-sym2}
\end{eqnarray}  
Both these symmetries also require $\mu_1^2=\mu_2^2$ and $m^2_{12}=0$. 
Note that in all the three naturally aligned scenarios, cf. (\ref{so5}), (\ref{other-sym1}) and (\ref{other-sym2}), $\tan\beta$ as given in (\ref{tanb}) `consistently' gives an {\em indefinite} answer 0/0. After   spontaneous  electro\-weak  symmetry   breaking,  symmetry~(i)
predicts two pseudo-Goldstone  bosons ($h$, $a$), whilst symmetry~(ii)
predicts only one pseudo-Goldstone boson, i.e.~the CP-even Higgs boson
$h$.     However,    a    non-zero    soft     SO(5)-breaking    mass
parameter~$m^2_{12}$ can be  introduced to render the pseudo-Goldstone
bosons  sufficiently massive, in  agreement with  present experimental
data, similar to the SO(5) case shown in Figure~\ref{fig1}.   
Even though  the 2HDM  scenarios based  on  the symmetries~(i)
and~(ii) may be analyzed in a  similar fashion, our focus here will be
on the simplest realization of the SM alignment, namely, the MS-2HDM based on  the SO(5) group.  Nevertheless,  the results
that we  will be deriving in  the present study are  quite generic and
could apply to the less symmetric cases (i) and (ii) above as well.

Before concluding this section, we would like to comment that no CP violation is possible in the MS-2HDM, be it spontaneously or explicitly, at least up to one-loop level. This is due to the fact that the Higgs potential (\ref{VSO5}) remains CP-invariant after the RG and one-loop threshold effects, even if a generic soft Z$_2$-breaking term ${\rm Im}(m^2_{12}e^{i\xi})\neq 0$ with an arbitrary CP-phase $\xi$ is present. In particular, as illustrated in Figure~\ref{fig1} (right panel), a non-zero $\lambda_{5,6,7}$ cannot be induced via RG effects, and this is true 
to any order in perturbation theory. On the other hand, one-loop threshold 
effects could induce non-zero $\lambda_{5,6,7}$ of the following form: 
\begin{eqnarray}
\lambda_5 \ \sim \ \frac{\lambda_3^2 (m^2_{12})^2}{16\pi^2|m^2_{12}|^2} \;, \qquad 
\lambda_{6} \ \sim \ \frac{\lambda_1\lambda_3 m^2_{12}}{16\pi^2|m^2_{12}|} 
\;, \qquad 
\lambda_{7} \ \sim \ \frac{\lambda_2\lambda_3 m^2_{12}}{16\pi^2|m^2_{12}|}
\; .
\label{thres}
\end{eqnarray}  
However, the Higgs potential still remains CP-invariant, due to the fulfillment of the following 
conditions~\cite{Pilaftsis:1999qt}:
\begin{eqnarray}
{\rm Im}(m^4_{12}\lambda_5^*) \ = \ {\rm Im}(m^2_{12}\lambda_6^*) \ = \ {\rm Im}(m^2_{12}\lambda_7^*) \ = \ 0 \; .
\end{eqnarray}
Therefore, there is no `CP-crisis' arising from large contributions to electric dipole moments in the MS-2HDM, unlike in the case of MSSM.

\section{Misalignment Predictions \label{sec:MP}}

As discussed in Section~\ref{sec:RGE},  a realistic Higgs spectrum can
be obtained by softly breaking  the maximal SO(5) symmetry of the 2HDM
potential   at  some   high   scale  $\mu_X$   by  considering   ${\rm
  Re}(m^2_{12})\neq 0$. As a consequence, there will be some deviation
from  the  alignment  limit  in  the low-energy  Higgs  spectrum.   By
requiring that the  mass and couplings of the  SM-like Higgs boson in our MS-2HDM 
are consistent with the latest Higgs  data from the 
LHC~\cite{Aad:2014aba,
  couplings1, couplings2}, we can derive predictions for the remaining
scalar spectrum and compare them with the existing (in)direct limits on the heavy
Higgs sector.  Our subsequent numerical  results are derived  for the
Type-II 2HDM  scenario, but the  analysis could be easily  extended to
other 2HDM scenarios.

\begin{figure}[t!]
\centering
\includegraphics[width=8cm]{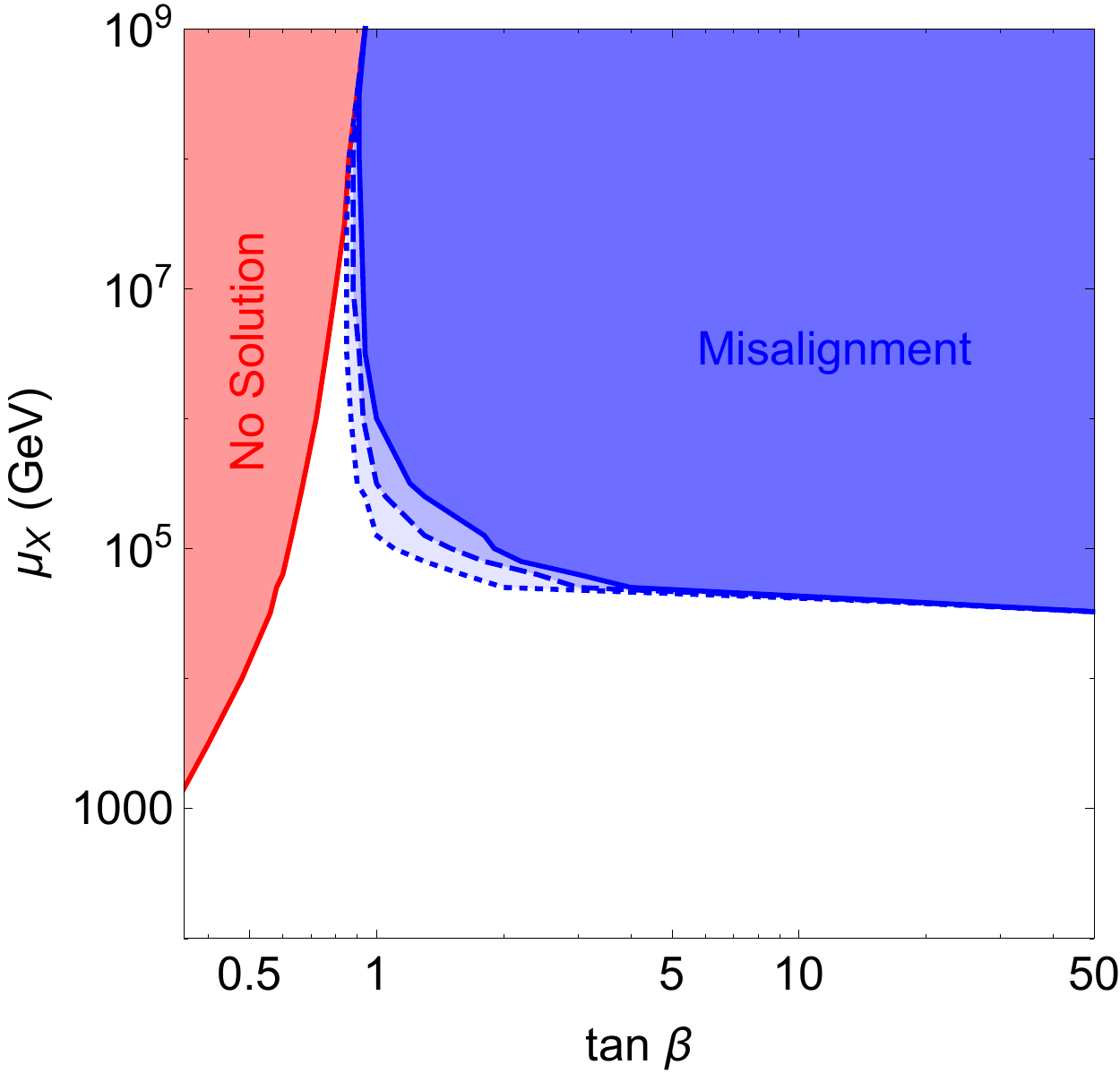}
\caption{Alignment constraints in the $(\tan\beta, \mu_X)$-plane of
  the maximally symmetric Type-II 2HDM. The blue shaded regions show the 
$1\sigma$ (dotted), $2\sigma$ (dashed) and $3\sigma$ (solid) exclusion regions 
from the alignment condition. The red shaded region is theoretically excluded 
in this model.
} \label{fig2} 
\end{figure}

For the  SM-like Higgs boson mass,  we will use  the $3\sigma$
allowed range from the recent CMS and ATLAS
Higgs    mass    measurements~\cite{couplings2, Aad:2014aba}: 
\begin{eqnarray}
M_H\ \in \ \big[124.1,~ 126.6\big]~{\rm GeV} \; . \label{MH}
\end{eqnarray}
For the Higgs  couplings to the SM vector bosons  and fermions, we use
the constraints in  the $(\tan\beta,~\beta-\alpha)$ plane derived from
a       recent       global       fit      for       the       Type-II
2HDM~\cite{Eberhardt:2013uba, eber2}.\footnote{Note  that in  our convention,
  the  couplings  of the  SM-like  Higgs  boson  to vector  bosons  is
  proportional to $\cos(\beta-\alpha)$ [cf.~(\ref{coup1})]. Hence, the
  natural  alignment limit  is obtained  for $\alpha=\beta$,  and {\it
    not}  for  $\alpha=\beta-\pi/2$,  as  conventionally used  in  
literature.}   For   a  given  set  of   SO(5)  boundary  conditions
$\big\{\mu_X,\tan\beta(\mu_X),\lambda(\mu_X)\big\}$, we  thus require that the
RG-evolved 2HDM  parameters at the  weak scale must satisfy  the above
constraints  on the  lightest  CP-even Higgs  boson sector.   This
requirement of {\it alignment} with the SM Higgs sector puts stringent
constraints   on   the   MS-2HDM   parameter  space,   as   shown   in
Figure~\ref{fig2}.   Here the  solid, dashed,  and dotted  blue shaded
regions  respectively  show   the  $1\sigma$, $2\sigma$  and  $3\sigma$
excluded regions due to misalignment of the mixing angle $\alpha$ from
its   allowed  range derived from the global fit.    
The  shaded   red  region   is  theoretically
inaccessible, as there is no viable solution to the RGEs in this area.
We  ensure  that  the  remaining  allowed (white) region  satisfies  the
necessary theoretical constraints,  i.e.~positivity and vacuum  stability of the
Higgs     potential,    and     perturbativity     of    the     Higgs
self-couplings~\cite{review}.   From Figure~\ref{fig2},  we  find that
there  exists  an upper  limit  of  $\mu_X\lesssim  10^9$ GeV  on  the
SO(5)-breaking  scale   of  the   2HDM  potential,  beyond   which  an
ultraviolet completion  of the theory must be  invoked.  Moreover, for
$10^5~{\rm GeV}\lesssim \mu_X \lesssim  10^9~{\rm GeV}$, only a narrow
range of $\tan\beta$ values are allowed.

\begin{figure}[t]
\centering
\includegraphics[width=8cm]{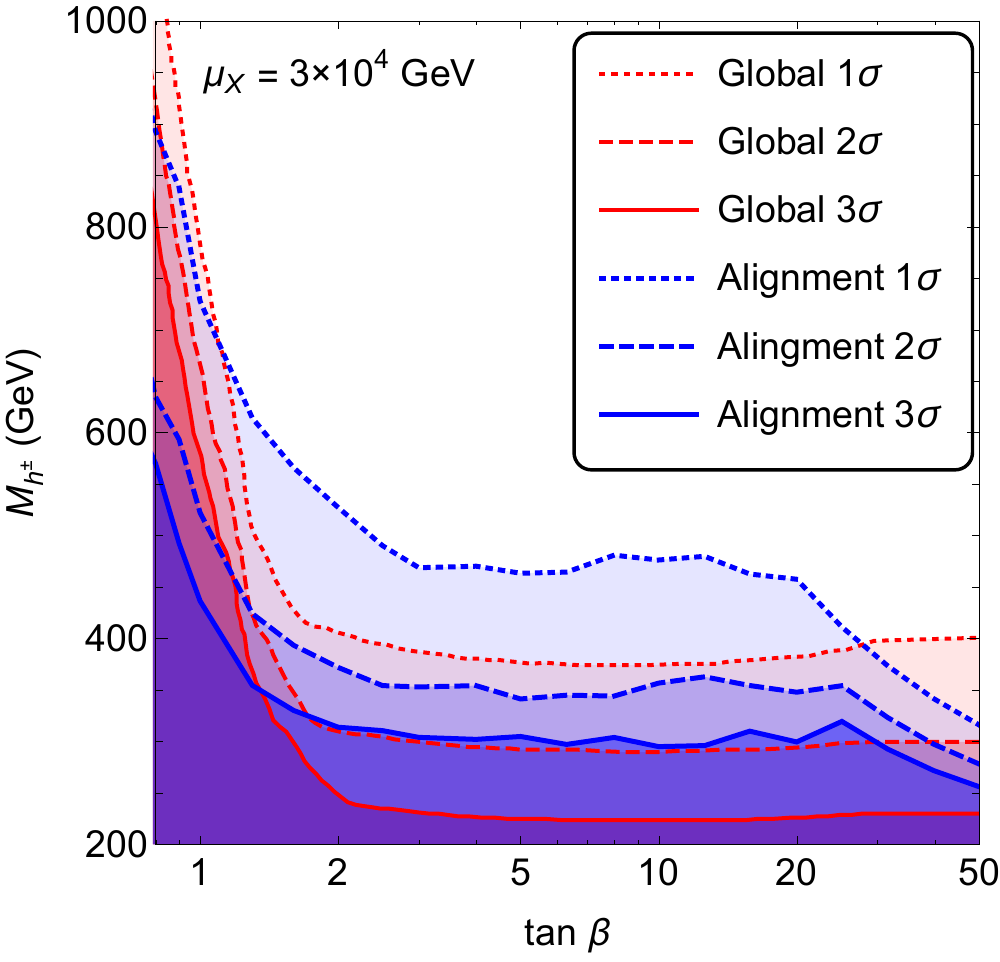}
\caption{The $1\sigma$ (dotted), $2\sigma$ (dashed) and $3\sigma$ (solid) lower limits 
on the charged Higgs mass obtained from the alignment condition (blue lines) 
in the maximally symmetric Type-II  2HDM with $\mu_X=3\times 10^{4}$ GeV. 
For comparison, the corresponding lower limits from a global fit 
are also shown (red lines). } \label{mhp}
\end{figure}

For  the  allowed   parameter  space  of  our  MS-2HDM   as  shown  in
Figure~\ref{fig2},  we obtain concrete  predictions for  the remaining
Higgs  spectrum.  In  particular,  the alignment  condition imposes  a
{\it lower} bound on the  soft breaking parameter Re$(m^2_{12})$, and hence,
on the heavy  Higgs spectrum. We compare this  limit with the existing
experimental limits on the  heavy Higgs sector of the 2HDM~\cite{PDG},
and find that the alignment  limits obtained here are more stringent in
a wide  range of  the parameter space.   The most  severe experimental
constraint comes from the charged  Higgs sector, which give significant 
contributions to various flavour observables, 
e.g. $B\to X_s\gamma$~\cite{flavor, flavor1, Hermann:2012fc}.  
For this, we use the
global fit results for the Type-II 2HDM from~\cite{Eberhardt:2013uba},
which includes limits derived from electroweak precision data, as well
as  flavour constraints  from  $\Delta m_{B_s}$  and $B\to  X_s\gamma$
relevant  for  the low  $\tan\beta$  region.  The  comparison of  the
existing  limit on  the  charged  Higgs-boson mass  as  a function  of
$\tan\beta$ with our predicted limits from the alignment condition for
a typical value of the boundary scale $\mu_X=3\times 10^{4}$ GeV 
is shown in Figure~\ref{mhp}.  It
is  clear that  the alignment  limits are  stronger than  the indirect
limits, except  in the very small  and very large $\tan\beta$ regimes. For 
$\tan\beta\lesssim 1$ region, the indirect limit obtained  from 
the $Z\to b\bar{b}$  precision observable becomes
the strictest~\cite{Deschamps:2009rh, Eberhardt:2013uba}. Similarly, for the 
large $\tan\beta\gtrsim 30$ case, the alignment limit can be easily obtained 
[cf.~(\ref{chat})] 
without requiring a large soft-breaking parameter $m_{12}^2$, and therefore, 
the lower limit on the charged Higgs mass derived from the misalignment condition 
becomes somewhat weaker in this regime. 

\begin{figure}[t]
\centering
\includegraphics[width=8cm]{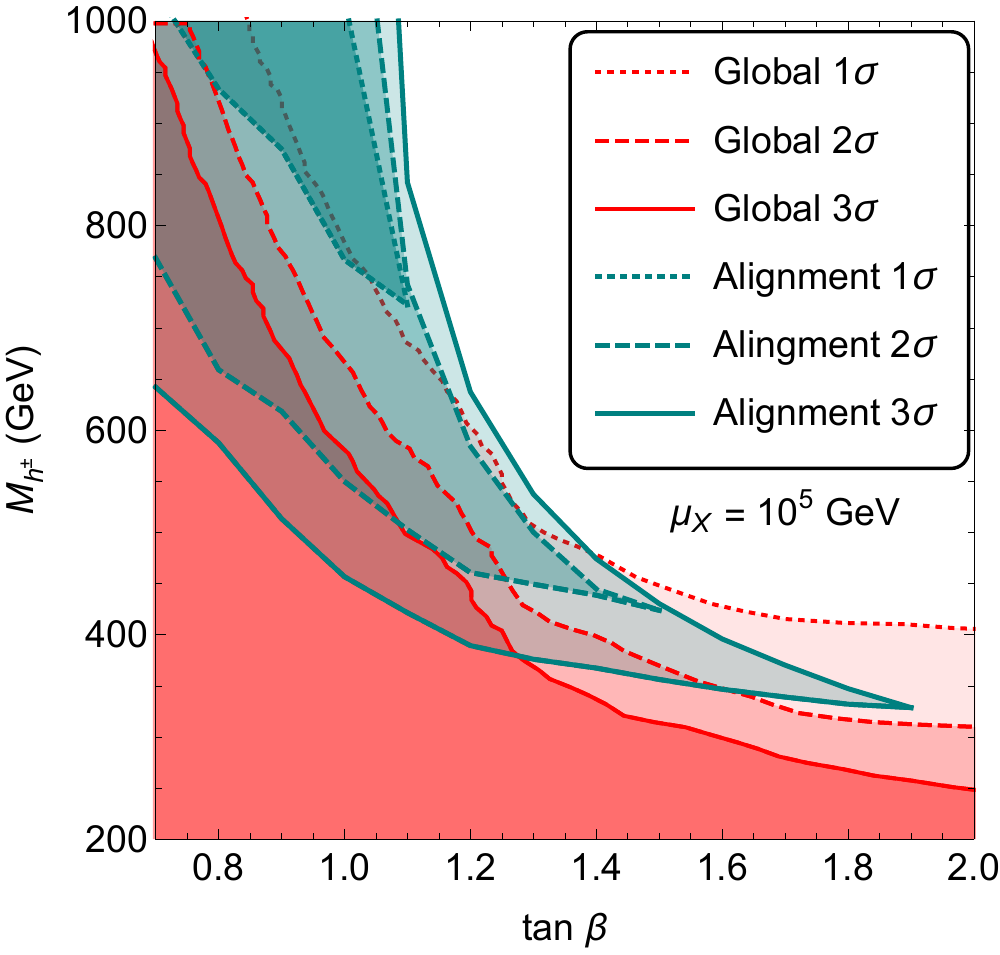}
\caption{Alignment  limits on the charged Higgs mass $M_{h^\pm}$ in the maximally 
symmetric Type-II  2HDM with $\mu_X=10^{5}$ GeV. The dark green regions show the 
$1\sigma$ (dotted), $2\sigma$ (dashed) and $3\sigma$ (solid) regions {\it allowed} by alignment constraints in the model. 
The existing lower limits from a global fit are also shown (red lines) for
  comparison.} \label{mhp2}
\end{figure}

From Figure~\ref{fig2}, it should be noted that for $\mu_X\gtrsim
10^5$~GeV, phenomenologically acceptable alignment is not possible in the MS-2HDM 
for large $\tan\beta$ {\it and} large $m^2_{12}$, while keeping the lightest CP-even Higgs boson
within  the experimentally  allowed range  (\ref{MH})  and maintaining
vacuum  stability up  to the  scale $\mu_X$.  Therefore, $\mu_X\gtrsim
10^5$~GeV  also  leads  to  an   {\it  upper}  bound  on  the  charged
Higgs-boson mass  $M_{h^\pm}$ from the misalignment condition, depending on $\tan\beta$. This is illustrated in Figure~\ref{mhp2} 
for $\mu_X=10^5$ GeV. Here the  green shaded regions show the $1\sigma$ (dotted), $2\sigma$ (dashed) and $3\sigma$ (solid) {\it allowed} regions, whereas the corresponding red shaded regions 
are the experimentally exclusion regions at $1\sigma$ (dotted), $2\sigma$ (dashed) and 
$3\sigma$ (solid).  On the other hand, for $\mu_X\lesssim  10^5$  GeV,  a phenomenologically acceptable aligned solution with an arbitrarily large $m^2_{12}$ is allowed for any value of $\tan\beta$ [cf.~Figure~\ref{fig2}],
and  hence  in  this  case,   there  exists  only  a  lower  limit  on
$M_{h^\pm}$, as  shown by the blue shaded  (exclusion) regions in Figure~\ref{mhp}.

Similar alignment constraints are obtained  for the heavy neutral pseudo-Goldstone
bosons $h$ and  $a$, which are predicted to be quasi-degenerate with the charged Higgs boson $h^\pm$ in the MS-2HDM [cf.~(\ref{mass-so5})]. The current experimental lower limits on the heavy neutral Higgs sector~\cite{PDG} are much weaker than the alignment constraints in this case. Thus, the MS-2HDM scenario provides a natural reason for the absence of a heavy Higgs signal below the top-quark threshold, and this has important consequences for the non-standard Higgs searches in the run-II phase of the LHC, as discussed in the following section.

\section{Collider Signals \label{sec:signals}}

In the alignment limit, the  couplings of the lightest CP-even Higgs
boson are exactly  similar to the SM Higgs  couplings, while the heavy
CP-even    Higgs    boson   preferentially    couples    to   fermions    (see
Appendix~\ref{app:spec}).   Therefore,  two   of  the  relevant  Higgs
production mechanisms at the LHC,  namely, the vector boson fusion and
Higgstrahlung processes are suppressed  for the gaugephobic heavy neutral Higgs sector. 
As a consequence, the only relevant production channels to probe the neutral Higgs
sector  of the  MS-2HDM  are the  gluon-gluon  fusion and  $t\bar{t}h$ 
($b\bar{b}h$) associated  production mechanisms at low (high) $\tan\beta$. 
For  the  charged Higgs sector of the MS-2HDM, the  dominant production 
mode is the associated production process: $gg\to \bar{t}bh^++t\bar{b}h^-$, irrespective of 
$\tan\beta$.

\subsection{Branching Fractions}
\begin{figure}[t]
\centering
\includegraphics[width=7.4cm]{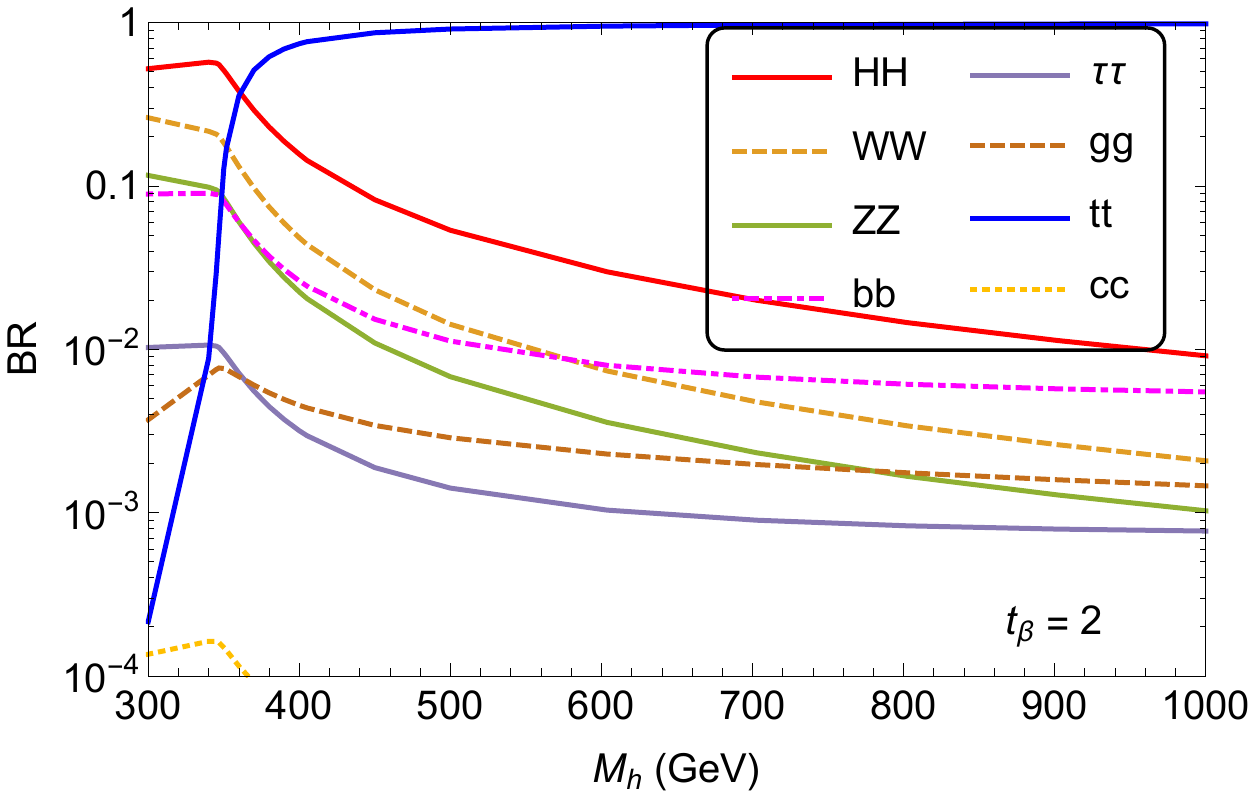}\\
\includegraphics[width=7.4cm]{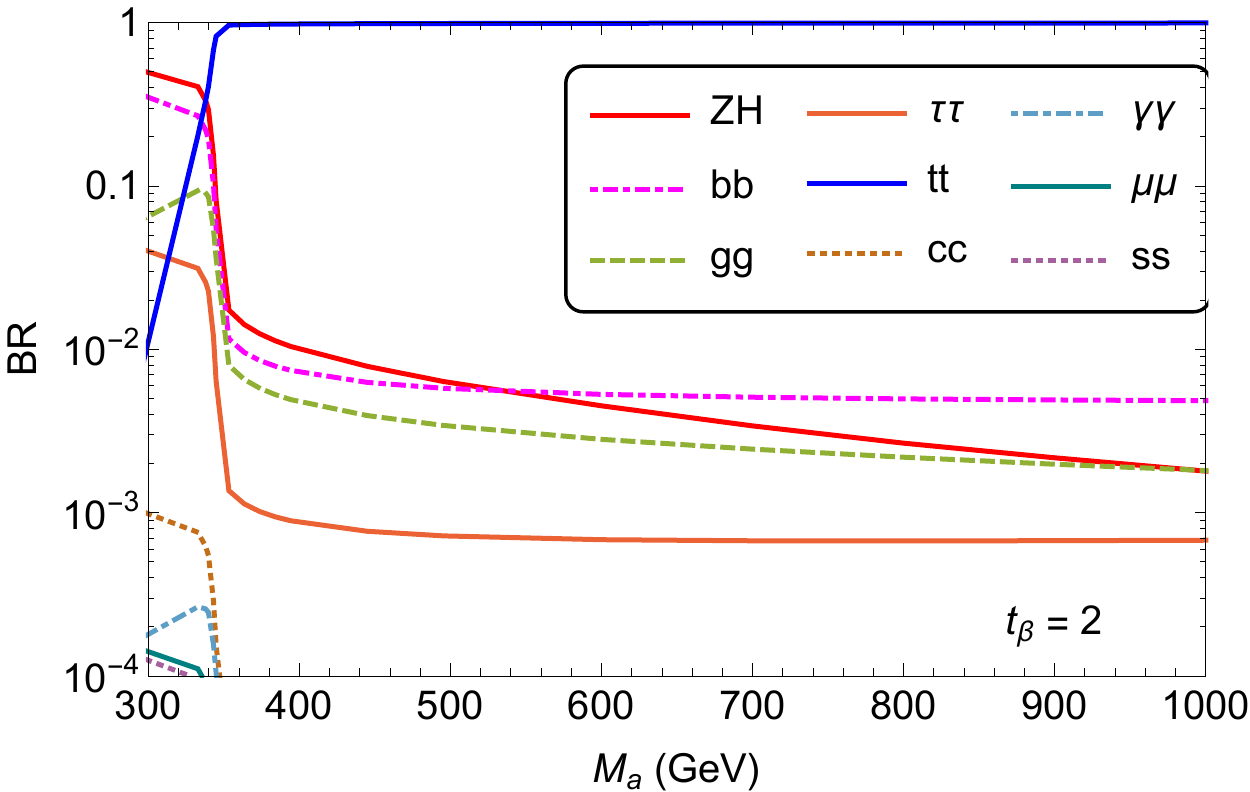}
\hspace{0.0cm}
\includegraphics[width=7.4cm]{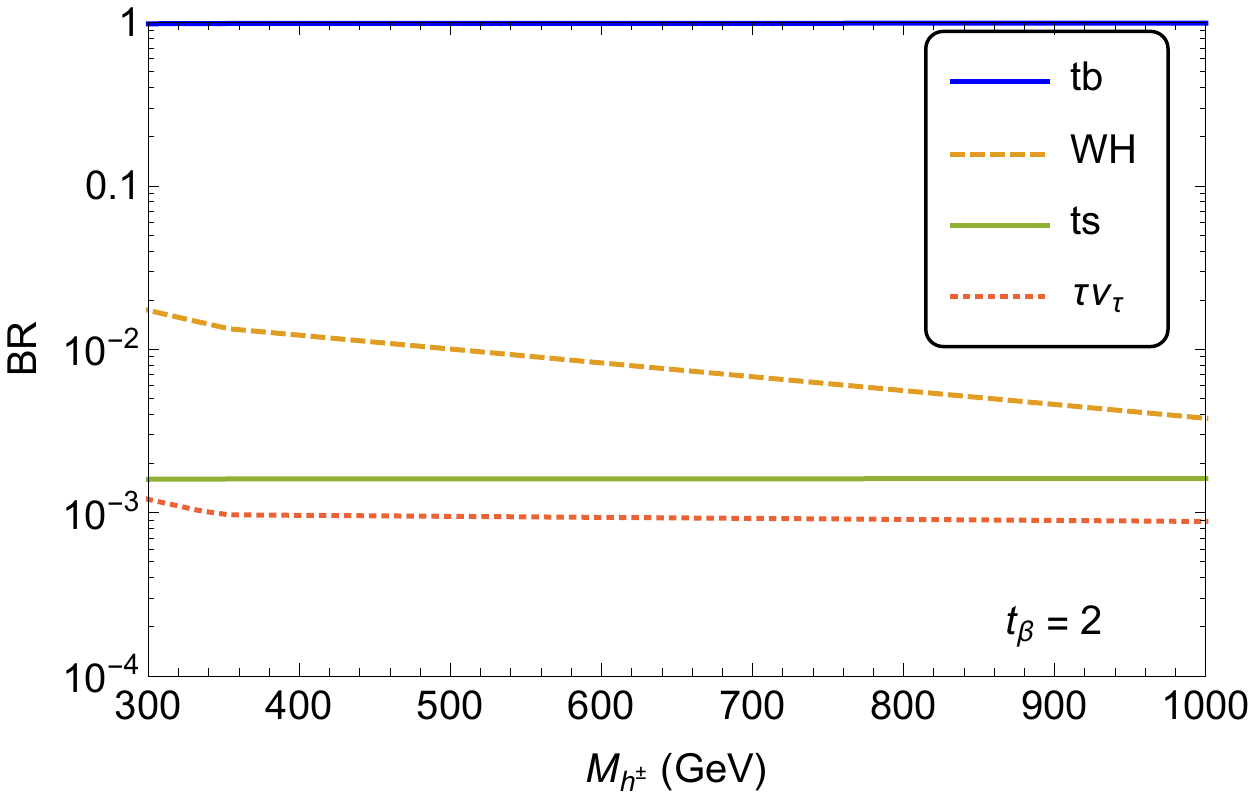}
\caption{The decay branching ratios of the heavy Higgs bosons
  in the maximally symmetric Type-II 2HDM for low $\tan\beta$.} \label{brlow}  
\end{figure}

\begin{figure}[t]
\centering
\includegraphics[width=10cm]{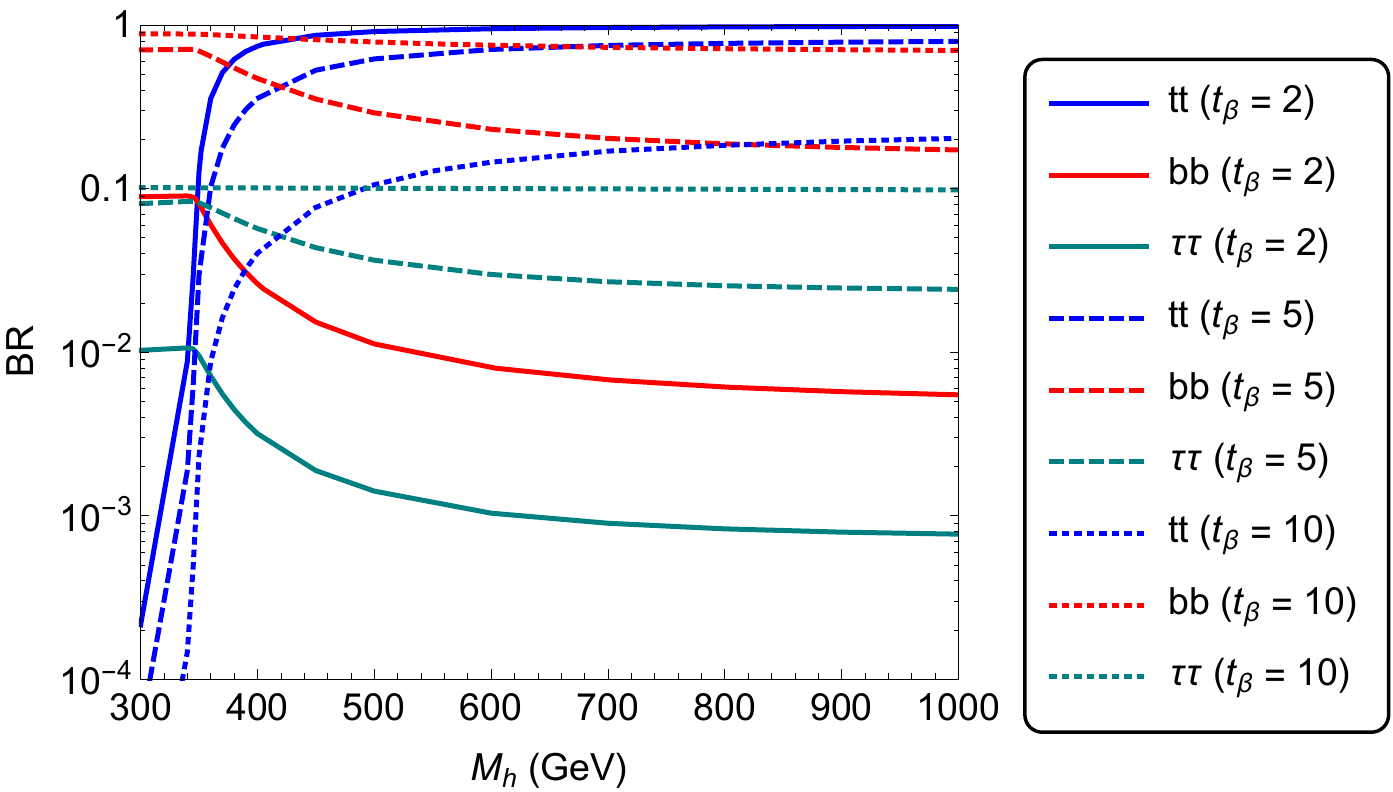}
\caption{Comparison of BR($h\to t\bar{t}$), BR($h\to b\bar{b}$) and BR($h\to \tau^+\tau^-$) for different values of $\tan\beta$ in the maximally symmetric Type-II 2HDM.} \label{brcomp}
\end{figure}

For our collider analysis, we calculate all the  
branching  ratios of the heavy Higgs sector in the MS-2HDM as  a function of  their masses 
using  the public {\tt C++} code {\tt 2HDMC}~\cite{2hdmc}. The results for $\tan\beta=2$ and with  
SO(5)-symmetric boundary conditions at $\mu_X=3\times 10^{4}$ GeV are  shown   in
Figure~\ref{brlow} for illustration. It  is clear that for the heavy neutral Higgs bosons, 
the $t\bar{t}$ decay mode is the  dominant one over most of  the MS-2HDM parameter space. 
However, this is true only for 
low $\tan\beta\lesssim  5$, since as we go to higher $\tan\beta$ values, 
the $b\bar{b}$ decay mode becomes dominant, with a sub-dominant  
contribution from $\tau^+\tau^-$, whereas the $t\bar{t}$ mode gets Yukawa suppressed. 
This is illustrated in Figure~\ref{brcomp}, where we compare BR($h\to t\bar{t}$), 
BR($h\to b\bar{b}$) and BR($h\to \tau^+\tau^-$) for three representative values of $\tan\beta=2$ (solid), 5 (dashed) and 
10 (dotted).  For the charged Higgs boson $h^{+(-)}$, the $t\bar{b}(\bar{t}b)$ mode is the 
dominant one over the entire parameter space, as shown in Figure~\ref{brlow} for $\tan\beta=2$, and this is true even for larger $\tan\beta$. 

\subsection{Charged Higgs Signal} \label{sec5.2}
The detection of a charged Higgs boson will be an unequivocal evidence for a beyond SM Higgs sector, and in particular, a `smoking gun' signal for a 2HDM. For $M_{h^\pm}<M_t$, i.e. below the top-quark threshold, stringent collider limits have been set on its production directly through top quark decays $t\to h^+ b$, followed by $h^+$ decays to $\tau^+\nu_\tau$~\cite{taunu, taunu_a}, $\tau^++$ jets~\cite{taujet} and $c\bar{s}$~\cite{cs, cs_a}. For charged Higgs boson masses above the top-quark threshold, the $h^+\to t\bar{b}$ decay channel opens up, and quickly becomes the dominant channel. In fact, in the Type-II 2HDM,   
the $h^+\bar{t}b$ coupling (\ref{coup-hp}) implies that for $M_{h^+}>M_t+M_b$, the  branching fraction  
of $h^+\to t\bar{b}$  is almost  100\% (cf.~Figure~\ref{brlow}), independent  of   $\tan\beta$. 
This leads  to  mostly $t\bar{t}b\bar{b}$ final states at the  LHC via
\begin{equation}
gg\ \to\ \bar{t} b h^+ + t\bar{b}h^- \ \to\ t\bar{t} b \bar{b}\;.
\label{ttbb}
\end{equation}
The experimental observation of this channel is challenging due to 
large  QCD backgrounds and the non-trivial event topology, involving at least four 
$b$-jets~\cite{hwg}. Nevertheless, we should emphasize here that (\ref{ttbb}) is the most promising channel for the charged Higgs signal in the MS-2HDM, because other interesting possibilities, such as $h^\pm \to aW^\pm, hW^\pm$~\cite{Coleppa:2014cca}, are not open in this scenario due to the kinematical constraints imposed by the quasi-degeneracy of the heavy Higgs sector [cf.~(\ref{mass-so5}) and Figure~\ref{fig1} (left panel)].   

\begin{figure}[t]
\centering
\includegraphics[width=8cm]{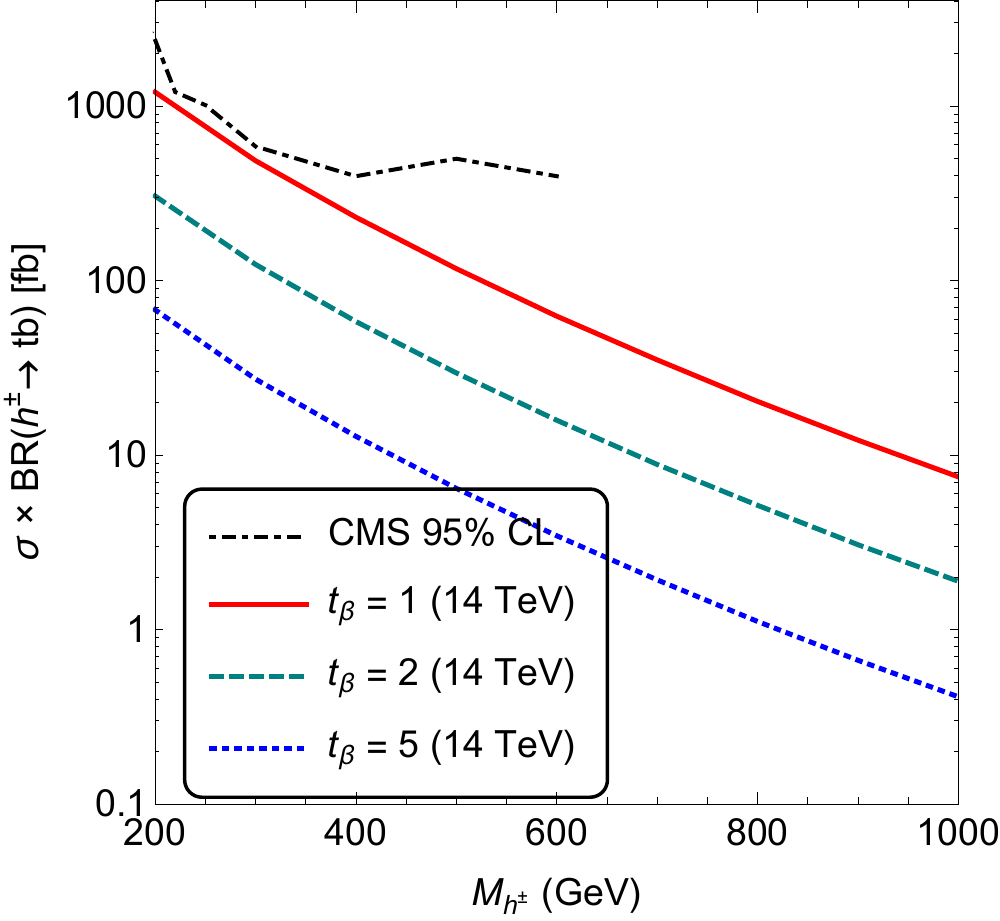}
\caption{Predictions for the cross section of the process (\ref{ttbb}) in the Type-II MS-2HDM 
at $\sqrt s=14$ TeV LHC for various values of $\tan\beta$. For comparison, we have also shown the current 95\% CL CMS upper limit from the $\sqrt s=8$ TeV data~\cite{tb}.}    
\label{ttbb-cross}
\end{figure}

A recent CMS study~\cite{tb} has presented for the first time a realistic analysis of the process (\ref{ttbb}), with the following decay chain: 
\begin{equation}
gg\ \to\ h^\pm tb \ \to \ (\ell \nu_\ell bb)(\ell'\nu_{\ell'}b)b 
\label{ttbb-ll}
\end{equation}
($\ell,\ell'$ beings electrons or muons). Using the $\sqrt s=8$ TeV LHC data, they have derived 95\% CL upper limits on the production cross section $\sigma(gg\to h^\pm tb)$ times the branching ratio BR($h^\pm\to tb$) as a function of the charged Higgs mass, as shown in Figure~\ref{ttbb-cross}. In the same Figure, we show the corresponding predictions at $\sqrt s=14$ TeV LHC in the Type-II MS-2HDM for some representative values of $\tan\beta$. The cross section predictions were obtained at leading order (LO) by implementing the 2HDM in {\tt MadGraph5}~\cite{mg5} and using the {\tt NNPDF2.3} PDF sets~\cite{nnpdf}.\footnote{For an updated and improved next-to-leading order (NLO) calculation, see~\cite{Flechl:2014wfa}.} A comparison of these cross sections with the CMS limit suggests that the run-II phase of the LHC might be able to probe a portion of the MS-2HDM parameter space using the process (\ref{ttbb}).  

In order to make a rough estimate of the $\sqrt s=14$ TeV LHC sensitivity to the charged Higgs signal (\ref{ttbb}) in the MS-2HDM, we perform a parton level simulation of the signal and background events using {\tt MadGraph5\_aMC@NLO}~\cite{mg5}. For the event reconstruction, we use 
some basic selection  cuts on  the transverse
momentum, pseudo-rapidity and dilepton invariant mass, following the recent CMS analysis~\cite{tb}:
\begin{eqnarray}
&& p_T^\ell \ > \ 20~{\rm GeV}, \quad |\eta^\ell| < 2.5, \quad \Delta R^{\ell\ell} > 0.4, \nonumber \\
&& M_{\ell\ell} > 12~{\rm GeV}, \quad |M_{\ell\ell}-M_Z| > 10~{\rm GeV},\nonumber \\
&& p_T^j \ > \ 30~{\rm GeV}, \quad |\eta^j| < 2.4, \quad \slashed{E}_T > 40~{\rm GeV}.
\label{cut}
\end{eqnarray}
Jets are reconstructed using the anti-$k_T$ clustering algorithm~\cite{anti-kT} with a distance parameter of 0.5. Since four $b$-jets are expected in the final state, at least two $b$-tagged jets are required in the signal events, and we assume the $b$-tagging efficiency for each of them to be  70\%.

\begin{figure}[t]
\centering
\includegraphics[width=7cm]{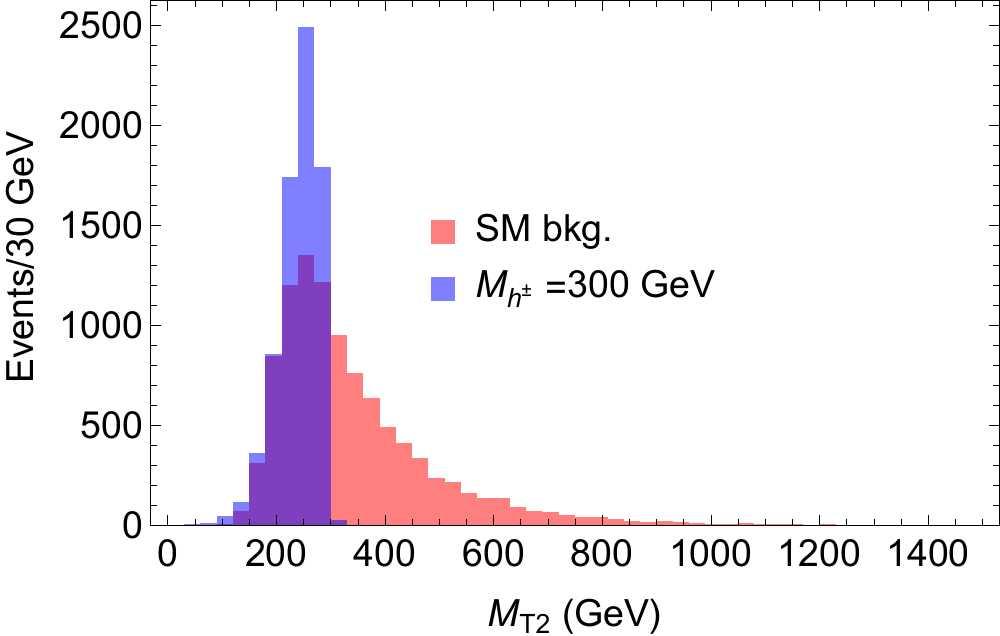}
\includegraphics[width=7cm]{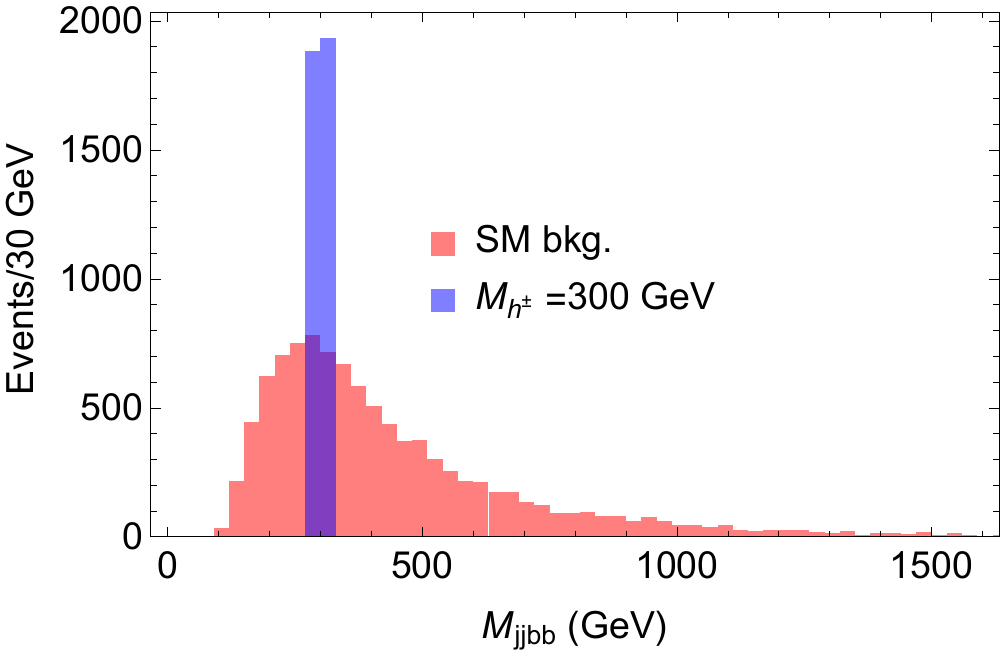}
\caption{An illustration of the charged Higgs boson mass reconstruction using the $M_{T2}$ (left panel) and invariant mass (right panel) observables. The irreducible SM background distribution is also shown for comparison.}    
\label{ttbb-dist}
\end{figure}

The inclusive SM  cross section for $pp\to t\bar{t}b\bar{b}+X$  is $\sim 18$ pb at NLO, with roughly 30\% uncertainty due to higher order QCD  corrections~\cite{pittau}. Most of the QCD background for the $4b+2\ell+\slashed{E}_T$ final state given by (\ref{ttbb-ll}) can be reduced significantly by reconstructing at least one top-quark. As we will show below, the remaining irreducible background due to SM $t\bar{t}b\bar{b}$ production can be suppressed with respect to the signal by reconstructing the charged Higgs boson mass, once a valid signal region is defined, e.g. in terms of an observed excess of events at the LHC in future.  For the semi-leptonic decay mode of top-quarks as in (\ref{ttbb-ll}), one cannot directly use an invariant mass observable to infer $M_{h^\pm}$, as both the neutrinos in the final state give rise to missing momentum. A useful quantity in this case is the $M_{T2}$ variable, also known as the `stransverse mass'~\cite{mt2}, defined as 
\begin{eqnarray}
M_{T2} \ = \ \underset{\left\{ \slashed{\mathbf p}_{T_{\rm a}}+\slashed{\mathbf p}_{T_{\rm b}}=\slashed{\mathbf p}_T\right\}}{\rm min}\Big[{\rm max}\left\{m_{T_{\rm a}},m_{T_{\rm b}}\right\}\Big] \;,
\label{mt2}
\end{eqnarray}
 where $\{{\rm a}\}, \{\rm b\}$ stand for the two sets of particles in the final state, each containing a neutrino with part of the missing transverse momentum ($\slashed{\mathbf p}_{T_{\rm {a,b}}}$). Minimization over all possible sums of these two momenta gives the observed missing transverse momentum $\slashed{\mathbf p}_T$, whose magnitude is the same as $\slashed{E}_T$ in our specific case. In (\ref{mt2}), $m_{T_{i}}$ (with $i=$a,b) is the usual transverse mass variable for the system $\{i\}$, defined as 
\begin{eqnarray}
m_{T_{i}}^2 \ = \ \left(\sum_{\rm visible} E_{T_i}+\slashed{E}_{T_i} \right)^2- \left(\sum_{\rm visible} {\mathbf p}_{T_i}+\slashed{\mathbf p}_{T_i} \right)^2 \; .
\end{eqnarray}
For the correct combination of the final state particles in (\ref{ttbb-ll}), i.e. for $\{{\rm a}\}=(\ell \nu_\ell bb)$ and $\{{\rm b}\}=(\ell' \nu_{\ell'}bb)$ in (\ref{mt2}), the maximum value of 
$M_{T2}$ represents the charged Higgs boson mass, with the $M_{T2}$ distribution smoothly dropping to zero at this point. This is illustrated in Figure~\ref{ttbb-dist} (left panel) for a typical choice of $M_{h^\pm}=300$ GeV. For comparison, we also show the $M_{T2}$ distribution for the SM background, which obviously does not have a sharp endpoint. Thus, for a given hypothesized signal region defined in terms of an excess due to $M_{h^\pm}$, we may impose an additional cut on $M_{T2}\leq M_{h^\pm}$ to enhance the signal (\ref{ttbb-ll}) over the irreducible SM background. 

Apart from the decay chain (\ref{ttbb-ll}) as considered in the CMS analysis~\cite{tb}, we also examine another decay chain involving hadronic decay modes of the secondary top-quark from the charged Higgs decay, i.e. 
\begin{eqnarray}
gg\ \to\ h^\pm tb \ \to \ (jjbb)(\ell \nu_{\ell}b)b \;.
\label{ttbb-jj}
\end{eqnarray}
In this case, the charged Higgs boson mass can be reconstructed using the invariant mass $M_{jjbb}$ for the correct combination of the $b$-quark jets. This is illustrated in Figure~\ref{ttbb-dist} (right panel) for $M_{h^\pm}=300$ GeV, along with the expected SM background. Thus, for the decay chain (\ref{ttbb-jj}), one can use an invariant mass cut of $M_{jjbb}$ around $M_{h^\pm}$ to observe the signal over the irreducible SM background. Note that the hadronic mode (\ref{ttbb-jj}) has a larger branching ratio, although from the experimental point of view, one has to deal with the uncertainties in the jet energy measurements, combinatorics and the resulting uncertainties in the invariant mass reconstruction of multiparticle final states.  

\begin{figure}[t]
\centering
\includegraphics[width=7cm]{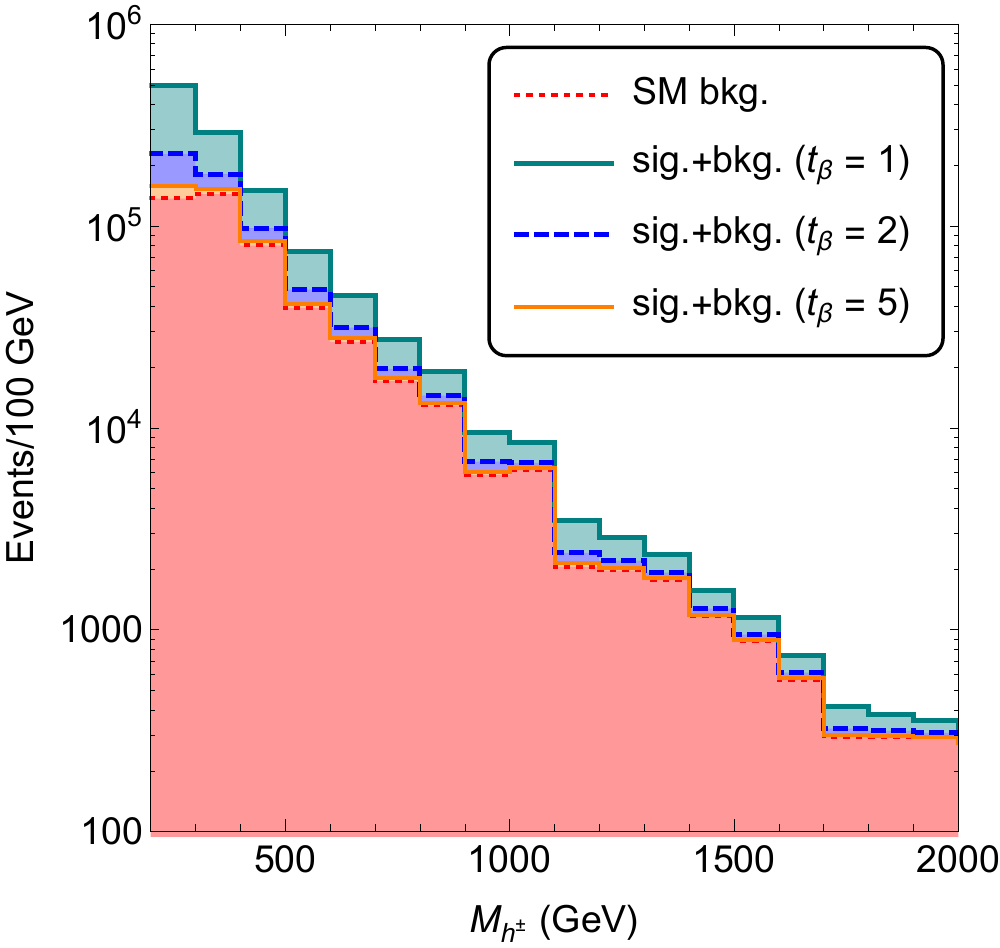}
\caption{Predicted number of events for the $t\bar{t}b\bar{b}$ signal
  from the charged pseudo-Goldstone boson in the MS-2HDM at $\sqrt
  s=14$ TeV LHC with $300~{\rm fb}^{-1}$ integrated luminosity. The
  results are shown for three different values of $\tan\beta=$1 (green
  solid), 2 (blue dashed) and 5 (orange solid).  The irreducible SM background (red dotted) is controlled by assuming an efficient mass reconstruction technique, as described in the text. } 
\label{2tb}
\end{figure}

Thus, in principle, we can obtain an observable charged Higgs signal in the MS-2HDM above the irreducible SM background by using  one of the methods shown in Figure~\ref{ttbb-dist} to reconstruct  efficiently the charged Higgs boson mass. Assuming this, we present an estimate of the signal to background ratio for the charged Higgs signal given by (\ref{ttbb}) at $\sqrt s=14$ TeV LHC with 300 fb$^{-1}$ for some typical values of $\tan\beta$ in Figure~\ref{2tb}. Since the mass of the charged Higgs boson is a priori unknown, we vary the charged Higgs mass, and for each value of $M_{h^\pm}$, we assume that it can be reconstructed around its actual value within 30 GeV uncertainty.\footnote{The uncertainty chosen here is always larger than the width of the charged Higgs boson $\Gamma_{h^\pm}$ in the entire mass range shown in Figure~\ref{2tb}. For instance, for $\tan\beta=2$, $\Gamma_{h^\pm} = 1.9$ GeV at $M_{h^\pm}=300$ GeV and $\Gamma_{h^\pm}=22$ GeV at $M_{h^{\pm}}=2$ TeV.} We believe such a mass resolution is feasible experimentally, given the fact that the top-quark mass resolution is of order of 1 GeV in both 
leptonic and hadronic channels~\cite{top-cms}.    
From Figure~\ref{2tb}, we see that the $t\bar{t}b\bar{b}$
channel (\ref{ttbb}) is effective for charged Higgs searches at the LHC for low $\tan\beta$ values. Note that the production
cross  section  $\sigma(gg\to  \bar{t}  b  h^+)$ decreases  rapidly  with
increasing $\tan\beta$ due to the Yukawa suppression [cf.~(\ref{coup-hp})], 
even though  BR($h^+\to
\bar{t}b$)  remains close to  100\%. 

\subsection{Heavy Neutral Higgs Signal} \label{sec5.3}

Since the heavy CP-even Higgs boson in the MS-2HDM is gaugephobic, most of the existing collider limits derived using the decay modes $h\to WW$~\cite{ww-cms, ww-cms2} and  $h\to ZZ$~\cite{zz-cms} do not apply in this case. The only existing searches relevant to the heavy CP-even sector of the MS-2HDM scenario are those based on $gg\to  h  \to  \tau^+\tau^-$  and  
$gg\to  b\bar{b}h  \to  b\bar{b} \tau^+\tau^-$~\cite{direct, direct2}.  However,  due  to the 
relatively small branching  ratio  of  $h\to  \tau^+\tau^-$, the  model-independent 
upper limits derived in~\cite{direct, direct2} are easily satisfied for the heavy Higgs spectrum presented here. Similarly, the $h\to \gamma\gamma$ branching ratio in the MS-2HDM is $\sim 10^2-10^3$ times smaller than that for the SM Higgs boson; therefore, the cross section limits derived from the $\gamma\gamma$ channel~\cite{gam-cms, gam-atlas} are also easily satisfied. So far there have been no direct searches for heavy neutral Higgs bosons involving $t\bar{t}$ and/or $b\bar{b}$ final states, mainly due to the challenges associated with uncertainties in the jet energy scales and the combinatorics arising from complicated multiparticle final states in a busy QCD environment. Nevertheless, these channels become pronounced in the MS-2HDM scenario, and hence, we will make here a preliminary attempt to study them.  

It is worth mentioning here that the Higgs pair production process $pp\to h\to HH$ (see e.g.~\cite{barger}) is another interesting possibility. However, as shown in (\ref{hHH}), the $h\to HH$ decay mode should also vanish in the exact alignment limit $\alpha\to \beta$, just like the $h\to VV$ decay modes. Therefore, the LHC limits derived using the $h\to HH$ channel~\cite{Khachatryan:2014jya, Aad:2014yja} are applicable only below the top threshold $M_h\leq 2M_t$ in the MS-2HDM (cf.~Figure~\ref{brlow}). On the other hand, the lower limits on the heavy Higgs sector, as derived in Section~\ref{sec:MP} (e.g.~Figures~\ref{mhp} and~\ref{mhp2}) strongly suggest a mass spectrum above the $t\bar{t}$ threshold, where the $h\to HH$ branching fraction drops orders of magnitude below that of $h\to t\bar{t}~(b\bar{b})$ at low (high) $\tan\beta$.

In light of the above discussion, we propose a new search channel for the heavy neutral Higgs boson in the MS-2HDM via the $t\bar{t}t\bar{t}$ final state: 
\begin{equation}
gg\ \to\  t\bar{t}h\  \to\ t\bar{t}t\bar{t}\; .
\label{4tp}
\end{equation}
Such four top final states have been proposed before in the context of
other  exotic  searches   at  the  LHC, e.g.~composite top~\cite{4topcomp1, 4topcomp2, 4topcomp3}, low-scale extra-dimensions~\cite{4topx1, 4topx2} and SUSY with light stops and gluinos~\cite{4topsusy}.  However,  their relevance for heavy Higgs searches have not been explored so far. We note here that the existing 95\% CL experimental upper limit on the four top production cross section is 59 fb from ATLAS~\cite{atlas-4t} and 32 fb from CMS~\cite{cms-4t}, whereas the SM prediction for the inclusive cross section of the process $pp\to t\bar{t}t\bar{t}+X$ is about 10-15 fb~\cite{Bevilacqua:2012em}. 

\begin{figure}[t]
\centering
\includegraphics[width=8cm]{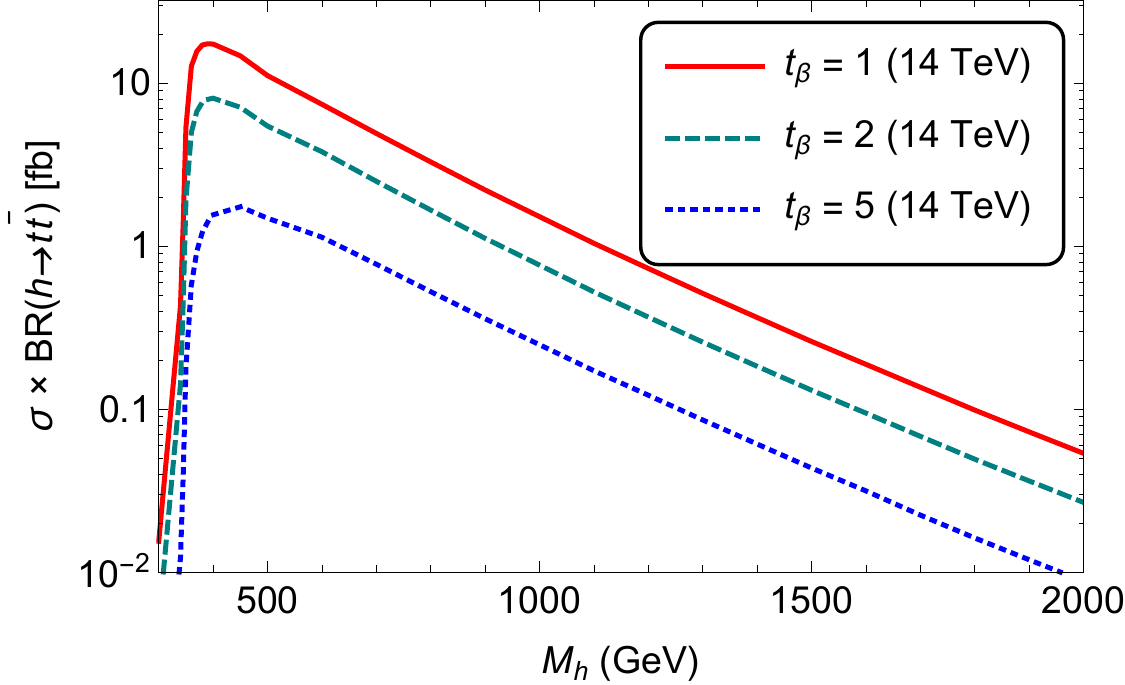}
\caption{Predictions for the cross section of the process (\ref{4tp}) in the Type-II MS-2HDM at $\sqrt s=14$ TeV LHC for various values of $\tan\beta$. } \label{4tx}
\end{figure}

To get a rough estimate of  the signal to background ratio for our new
four-top signal,  we perform a  parton-level simulation of  the signal
and   background  events   at  LO  in   QCD   using  {\tt
  MadGraph5\_aMC@NLO}~\cite{mg5}   with   {\tt NNPDF2.3}  PDF sets~\cite{nnpdf}.  
For the  inclusive  SM  cross  section  for  the
four-top final  state at $\sqrt s=14$ TeV LHC, we obtain 11.85 fb, whereas  our proposed four-top 
signal  cross sections are
found   to  be   comparable  or   smaller  depending   on   $M_h$  and
$\tan\beta$, as shown in Figure~\ref{4tx}. However, since  we expect  one of  the  $t\bar{t}$ pairs
coming from  an on-shell  $h$ decay to  have an invariant  mass around
$M_h$,  we can  use this information to significantly boost the signal over the irreducible SM background. Note that all the predicted cross sections shown in Figure~\ref{4tx} are well below 
the current  experimental  upper bound~\cite{cms-4t}. 

Depending on the $W$ decay mode from $t\to Wb$, there are 35 final states for four top decays. According to a recent ATLAS analysis~\cite{thesis}, the experimentally favoured channel is the semi-leptonic/hadronic final state with two same-sign isolated leptons. Although the branching fraction for this topology (4.19\%) is smaller than most of the other channels, the presence of two same-sign leptons in the final state allows us to reduce the large QCD background substantially, including that due to the SM production of $t\bar{t}b\bar{b}+$jets.\footnote{For a detailed analysis of the reducible and irreducible four top background, see~\cite{thesis}.} Therefore, we will only consider the following decay chain in our preliminary analysis: 
\begin{eqnarray}
gg\ \to\  t\bar{t}h\  \to\ (t\bar{t})(t\bar{t}) \ \to \  
\Big( (\ell^\pm \nu_{\ell}b)(jjb)\Big)\Big((\ell'^\pm \nu_{\ell'}b)(jjb)\Big)\; .
\label{4t-ll}
\end{eqnarray}
For event reconstruction, we will use the same selection cuts as in (\ref{cut}), and in addition, following~\cite{thesis}, 
we require the scalar sum of the $p_T$ of all leptons and jets (defined as $H_T$) to exceed 350  GeV.    

\begin{figure}[t]
\centering
\includegraphics[width=7cm]{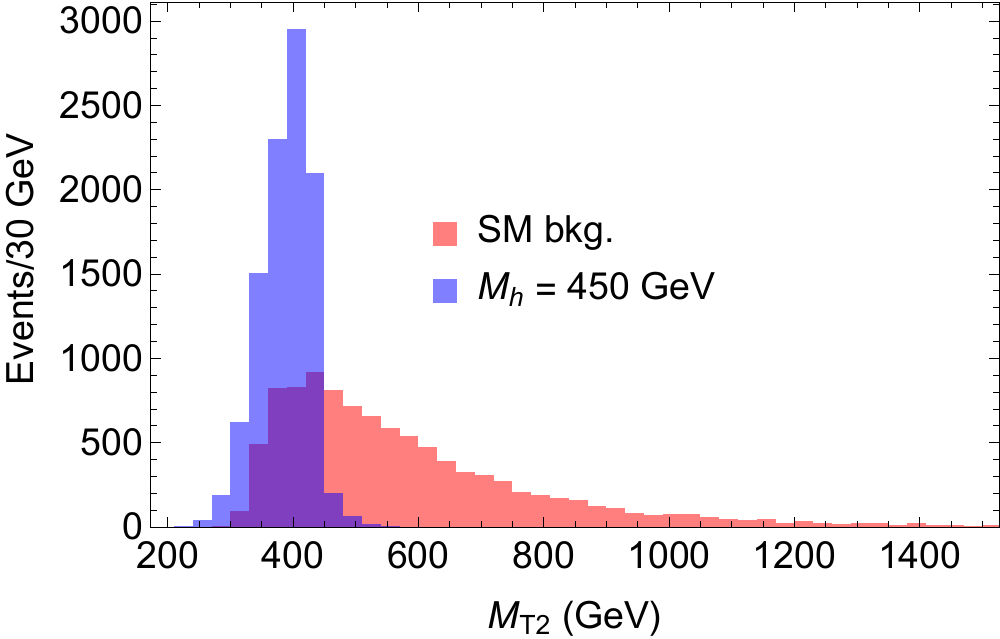}
\caption{An illustration of the heavy CP-even Higgs boson mass reconstruction using $M_{T2}$ observable. The irreducible SM background distribution is also shown for comparison.}    
\label{mt2-4t}
\end{figure}

As in the charged Higgs boson case [cf.~(\ref{ttbb-ll})], the heavy Higgs mass can be reconstructed from the signal given by (\ref{4t-ll}) using the $M_{T2}$ endpoint technique. The correct combination of visible final states in (\ref{4t-ll}) will lead to a smooth drop at $M_h$ in the $M_{T2}$ distribution, as illustrated in Figure~\ref{mt2-4t} for a typical choice of $M_h=450$ GeV. As shown in the same figure, the SM background does not exhibit such a feature, and hence, an additional selection cut on $M_{T2}\leq M_h$ can be used to enhance the signal to background ratio in the signal region. 

\begin{figure}[t]
\centering
\includegraphics[width=7cm]{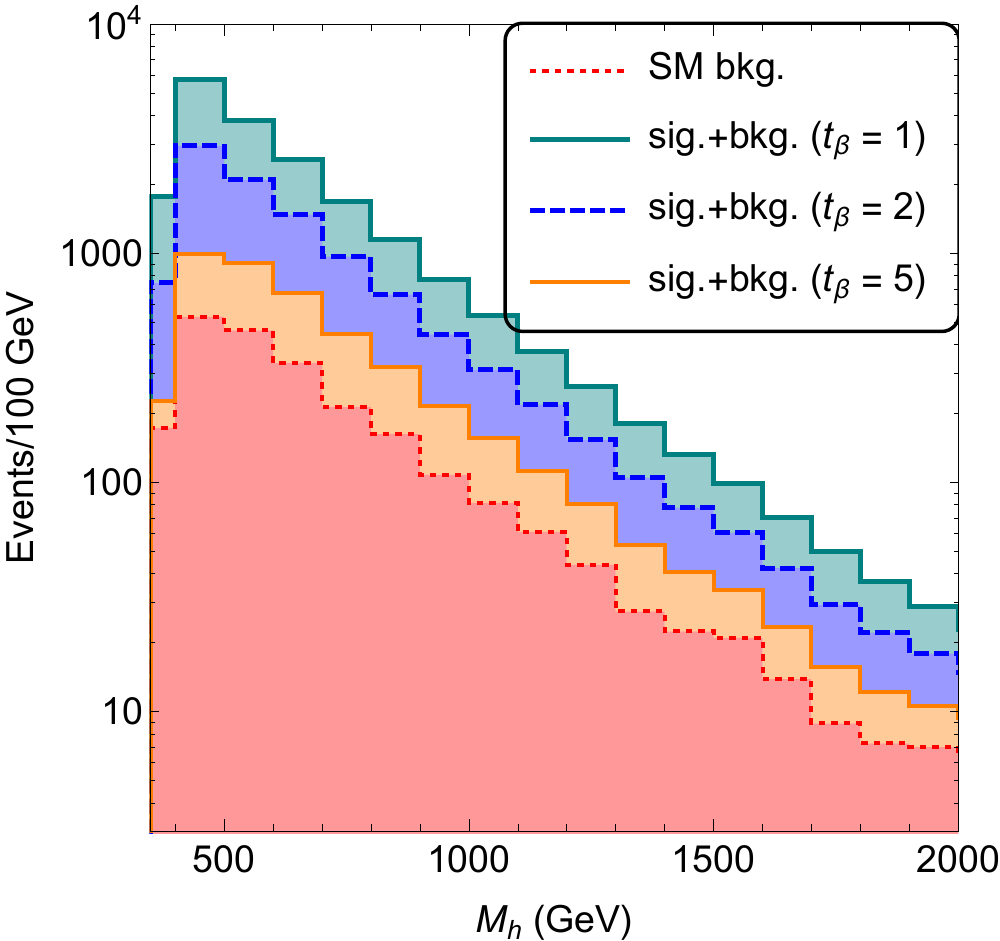}
\caption{Predicted number of  events for the $t\bar{t}t\bar{t}$ signal
  from  the neutral pseudo-Goldstone  boson in  the MS-2HDM  at $\sqrt
  s=14$ TeV  LHC with  $300~{\rm fb}^{-1}$ integrated  luminosity. The
  results are shown for three different values of $\tan\beta=$1 (green
  solid), 2 (blue  dashed) and 5 (orange solid).  The SM background (red dotted) is controlled by assuming an efficient mass reconstruction technique, as outlined in the text. } \label{4t}
\end{figure}

Our simulation results  for the predicted number of signal and background events for the process (\ref{4t-ll}) at $\sqrt s=14$ TeV LHC with 300 fb$^{-1}$ luminosity are  shown in Figure~\ref{4t}. The signal events are shown for  three representative
values of $\tan\beta$. Here we vary the a priori unknown heavy Higgs mass, and for each value of $M_{h}$, we assume that it can be reconstructed around its actual value within 30 GeV uncertainty, which  should be feasible experimentally, as argued at the end of Subsection~\ref{sec5.2}. From this preliminary
analysis, we  find that the
$t\bar{t}t\bar{t}$ channel provides the most promising collider signal
to  probe the  heavy Higgs  sector in  the MS-2HDM  at low  values of
$\tan\beta \lesssim  5$.

The above analysis  is also applicable for the  CP-odd Higgs boson
$a$,  which  has  similar  production  cross  sections  and  $t\bar{t}$
branching fractions as the CP-even Higgs $h$. However, 
the $t\bar{t}h(a)$ production cross section  as well as the $h(a)\to t\bar t$ branching
ratio decreases with  increasing $\tan\beta$. This is due  to the fact
that   the   $ht\bar{t}$   coupling   in  the   alignment   limit   is
$\cos\alpha/\sin\beta\sim \cot\beta$, which is same as the $at\bar{t}$
coupling [cf.~(\ref{couplings})]. Thus,  the high $\tan\beta$ region of
the MS-2HDM  cannot be searched  via the $t\bar{t}t\bar{t}$ channel  proposed above,
and one  needs to consider  the channels involving  down-sector Yukawa
couplings, e.g. $b\bar{b}b\bar{b}$ and $b\bar{b}\tau^+\tau^-$, which
are very challenging in  the LHC environment~\cite{hwg}. For instance,
the SM  $b\bar{b}b\bar{b}$ cross  section at $\sqrt  s=14$ TeV  LHC is
about  140  pb at  NLO~\cite{Bevilacqua:2013taa},  whereas the  $pp\to
b\bar{b} h\to b\bar{b}b\bar{b}$ signal cross section for $M_h=300$ GeV
and $\tan\beta=10$  is only  about 0.3 pb  at NLO, as  estimated using 
the public {\tt FORTRAN} code {\tt   SusHi}~\cite{sushi}.   In  practice,   one   would  require   a
sophisticated  jet   substructure  technique~\cite{seymour,  BDRS}  to
disentangle such a tiny signal from the huge QCD background. It is also worth commenting here that the simpler process  $pp\to h\to t\bar{t}~(b\bar{b})$ at low (high) $\tan\beta$ similarly suffers from a huge SM $t\bar{t}$ ($b\bar{b}$) QCD background, even after imposing an $M_{t\bar{t}~(b\bar{b})}$ cut. 


Before concluding this section, we should clarify that although we were able to obtain a sizeable signal-to-background ratio in the low $\tan\beta$ regime for the signals (\ref{ttbb-ll}), (\ref{ttbb-jj}) and (\ref{4t-ll})  using efficient mass reconstruction techniques in the signal region, as described in Subsections~\ref{sec5.2} and \ref{sec5.3} above, these results are valid only at the parton level. In a realistic detector environment, the sharp features of the signal shown in Figures~\ref{ttbb-dist} and \ref{mt2-4t}  may not survive, and therefore, the signal-to-background ratio might get somewhat reduced than that shown here. However, a detailed realistic detector-level  analysis  of  these  signals in our MS-2HDM scenario,  including realistic  top reconstruction  efficiencies and smearing effects, is beyond the scope and main focus of this article, and is being  pursued in  a separate dedicated study.

\section{Conclusions \label{sec:concl}}

We have analyzed  the symmetries of the 2HDM scalar potential 
to naturally justify the so-called SM alignment limit, independently of
the  heavy  Higgs  spectrum  and  the  value  of  $\tan\beta$.  
We show that in the Type-II 2HDM, there exist {\em only} three different symmetry realizations, cf. (\ref{so5}), (\ref{other-sym1}) and (\ref{other-sym2}), which could lead to a natural alignment by satisfying (\ref{align-gen}) for {\em any} value of $\tan\beta$. In the context of the Maximally Symmetric Two
Higgs  Doublet Model  based  on the  SO(5)  group,  we demonstrate how small
deviations from this alignment limit are naturally induced by RG 
effects  due  to  the   hypercharge  gauge  coupling  $g'$  and  third
generation Yukawa  couplings, which also break  the custodial symmetry
of  the theory.   In  addition, a  non-zero  soft SO(5)-breaking  mass
parameter is required to yield a viable Higgs spectrum consistent with
the  existing experimental constraints.   Employing the  current Higgs
signal strength  data from the  LHC, which disfavour  large deviations
from the alignment limit, we  derive important constraints on the 2HDM
parameter space.  In particular, we  predict lower limits on the heavy
Higgs spectrum,  which prevail the present  limits in a  wide range of
parameter space.   Depending on the  scale where the  maximal symmetry
could be  realized in  nature, we  also obtain an  upper limit  on the
heavy Higgs masses in certain  cases, which could be completely probed
during  the  run-II phase  of  the LHC.   Finally,  we  propose a  new
collider signal  with four  top quarks in  the final state,  which can
become a valuable observational tool to directly probe the heavy Higgs
sector of the 2HDM in the alignment limit for low values of $\tan\beta$. 
It would be interesting to investigate how this tool could be applied 
to supersymmetric theories in the alignment limit. 


\subsection*{Acknowledgments}
A.P.   thanks Celso  Nishi for  a clarifying  remark that  led  us to
include the fourth  footnote  in this article.  P.S.B.D. thanks Otto Eberhardt 
and  Martin Wiebusch for helpful discussions regarding their global fit results 
in~\cite{Eberhardt:2013uba, eber2}. We also thank Rohini Godbole, Romain Madar, 
Yvonne Peters and Jos\'{e} Santiago for valuable comments on the four top analysis. 
The work  of P.S.B.D. and   A.P.   is   supported   by  the   Lancaster-Manchester-Sheffield
Consortium for Fundamental Physics under STFC grant ST/J000418/1.


\appendix

\section{Higgs Spectrum and Couplings in a General 2HDM}\label{app:spec}

Here  we will  restrict our  discussion to  2HDM  potentials realizing
CP-conserving  vacua.    In  this   case,  the  minimization   of  a
CP-conserving 2HDM  potential (\ref{pot}) yields  the following real
non-negative vacuum expectation values (VEVs):
\begin{eqnarray}
\langle \Phi_1 \rangle \ = \ \frac{1}{\sqrt 2} \left(\begin{array}{c} 0
  \\ v_1 \end{array}\right),  
\qquad  
\langle \Phi_2 \rangle \ = \ \frac{1}{\sqrt 2} \left(\begin{array}{c} 0
  \\ v_2 \end{array}\right), 
\end{eqnarray}
where $v=\sqrt{v_1^2+v_2^2}=246.2$~GeV  is the SM  electroweak VEV and
for later convenience, we  define $\tan\beta \equiv v_2/v_1$.  The two
scalar doublets can  be expanded in terms of  eight real scalar fields
as follows:
\begin{eqnarray}
\Phi_j \ = \ \left(\begin{array}{c} \phi_j^+ \\ \frac{1}{\sqrt
    2}(v_j+\phi_j+ia_j) \end{array} \right)\; , 
\label{expand-phi}
\end{eqnarray}
with $j = 1,2$.  After  spontaneous symmetry breaking, there are three
Goldstone   modes  ($G^\pm,G^0$),   which   become  the   longitudinal
components  of the SM  $W^\pm$ and  $Z$ bosons.  Thus, there  are five
remaining physical scalar mass eigenstates: two CP-even ($h,H$), one
CP-odd ($a$)  and two charged  ($h^\pm$) scalars. The mixing  in the
CP-odd  and charged  sectors is  governed by the angle $\beta$ defined above:  
\begin{align}
\left(\begin{array}{c} G^\pm \\ h^\pm \end{array}\right) \ & =  \ 
\left(\begin{array}{cc} c_\beta & s_\beta \\ -s_\beta & c_\beta \end{array}\right) 
\left(\begin{array}{c} \phi_1^\pm \\ \phi_2^\pm \end{array}\right) \;
, \nonumber \\  
\left(\begin{array}{c} G^0 \\ a \end{array}\right) \ & = \ 
\left(\begin{array}{cc} c_\beta & s_\beta \\ -s_\beta & c_\beta \end{array}\right) 
\left(\begin{array}{c} a_1 \\ a_2 \end{array}\right) \; , 
\end{align}
where $c_\beta\equiv \cos\beta$,  $s_\beta \equiv \sin\beta$. On the other hand, in the
CP-even sector, we have a new mixing angle $\alpha$:
\begin{align}
\left(\begin{array}{c} H \\ h \end{array}\right) \ & = \ 
\left(\begin{array}{cc} c_\alpha & s_\alpha \\ -s_\alpha &
  c_\alpha \end{array}\right)  
\left(\begin{array}{c} \phi_1 \\ \phi_2 \end{array}\right) \; .
\end{align}
The
corresponding     physical      mass     eigenvalues     are     given
by~\cite{Haber:1993an,Pilaftsis:1999qt}
\begin{align}
M^2_{h^\pm} \ & = \ \frac{m_{12}^2}{s_\beta
  c_\beta}-\frac{v^2}{2}\left( \lambda_4+ \lambda_5\right)
 +\frac{v^2}{2s_\beta c_\beta}\left( \lambda_6
c_\beta^2+ \lambda_7 s_\beta^2\right),
\nonumber \\ 
M_a^2 \ &  = \  M^2_{h^\pm}+\frac{v^2}{2}\left( \lambda_4 - 
\lambda_5\right), \nonumber\\ 
M_H^2 \ & = \ \frac{1}{2}\left[(A+B)-\sqrt{(A-B)^2+4C^2}\right], \nonumber\\
M_h^2 \ & = \ \frac{1}{2}\left[(A+B)+\sqrt{(A-B)^2+4C^2}\right], \label{mass}
\end{align} 
where we have defined $\tan{2\alpha} = 2C/(A-B)$, and 
\begin{align}
A \ & = \ M_a^2s_\beta^2+v^2\left( 2\lambda_1 c_\beta^2+ 
\lambda_5 s^2_\beta+2 \lambda_6 s_\beta
  c_\beta\right),\nonumber \\ 
B \ & = \ M_a^2c_\beta^2+v^2\left( 2\lambda_2 s_\beta^2+ 
\lambda_5 c^2_\beta + 2\lambda_7 s_\beta c_\beta
  \right) , \qquad \\ 
C \ & = \ -M_a^2 s_\beta c_\beta + v^2\left( \lambda_{34}s_\beta c_\beta
  + \lambda_6 c^2_\beta + \lambda_7 s^2_\beta
  \right) . \nonumber  
\end{align}
with $\lambda_{34}=\lambda_3+\lambda_4$. The SM Higgs field is given by 
\begin{eqnarray}
H_{\rm SM} \ = \  \phi_1 \cos\beta + \phi_2 \sin \beta \ = \ 
H\cos(\beta-\alpha)+h\sin(\beta-\alpha) \; . 
\label{HSM}
\end{eqnarray}

From~(\ref{HSM}), the  couplings of  $h$ and $H$  to the  gauge bosons
($V=W^\pm, Z$) with  respect to the SM Higgs  couplings $g_{H_{\rm SM}VV}$
are given by
\begin{eqnarray}
g_{hVV} \ = \ s_{\beta-\alpha} \; , \qquad g_{HVV} \ = \ c_{\beta-\alpha} \; . 
\label{coup1}
\end{eqnarray}
Similarly,  unitarity  constraints  uniquely fix  the other  Higgs-Higgs-$V$ 
couplings~\cite{hunter}:
\begin{eqnarray}
&& g_{haZ} \ = \ \frac{g}{2\cos\theta_w}c_{\beta-\alpha} \; , \qquad
  g_{HaZ} \ = \ \frac{g}{2\cos\theta_w}s_{\beta-\alpha} \; ,
  \nonumber\\ 
&&  g_{h^+hW^-} \ = \ \frac{g}{2}c_{\beta-\alpha} \; , \quad
  g_{h^+HW^-} \ = \ \frac{g}{2}s_{\beta-\alpha} \; , \quad 
g_{ah^\pm W^\mp} \ = \ \frac{g}{2} \; ,
\label{coup2}
\end{eqnarray} 
(where $\theta_w$ is the weak mixing angle) in order to satisfy the sum rules~\cite{sum}
\begin{align}
& g^2_{haZ}+g^2_{HaZ} \  = \  \frac{1}{\cos^2\theta_w}(g^2_{h^+hW^-}+g^2_{h^+HW^-}) \ = \ \frac{1}{4M_Z^2}g^2_{H_{\rm SM}ZZ} \; , \\
& g^2_{h^+hW^-}+g^2_{h^+HW^-} \  = \ g^2_{ah^\pm W^\mp} \ = \ \frac{1}{4M_W^2}g^2_{H_{\rm SM}W^+W^-} \; . 
\end{align}

For our subsequent discussion, we also write down the $h$-$H$-$H$ coupling~\cite{Gunion:2002zf}: 
\begin{eqnarray}
g_{hHH} \ &  = & \ \frac{s_{\beta-\alpha}}{4v s_\beta c_\beta}
\Big[ 2(M_h^2+2M_H^2)s_{2\alpha}
 -\: 2(M_a^2+v^2\lambda_5)(s_{2\beta}+3s_{2\alpha})
  \nonumber \\ 
&& \qquad \qquad -\: v^2(\lambda_6-\lambda_7)(c_{2\beta}+3c_{2\alpha}) 
-\: v^2(\lambda_6+\lambda_7)(1+3c_{2(\beta-\alpha)})\Big]\; .
\label{hHH}
\end{eqnarray} 
Note that the coupling $g_{hHH}$ is proportional to $s_{\beta-\alpha}$
and so vanishes identically in the alignment limit $\alpha \to \beta$.

To obtain  a phenomenologically acceptable  theory, we need  to forbid
Higgs interactions  with tree-level  FCNCs.  This can  be accomplished
minimally  by imposing  appropriate discrete  ${\rm  Z}_2$ symmetries,
which will  explicitly break  in general the custodial
symmetries of the theory.  By  convention, we may take $u_R$ to couple
to $\Phi_2$,~i.e.~$h_1^u=0$, and then $\Phi_1$  ($\Phi_2$) to couple
to  $d_R$,   with  $h_2^d=0$   ($h_1^d=0$),  in  a   Type-II  (Type-I)
realization of the  2HDM.  As our interest is in  the Type-II 2HDM, we
only list the  Yukawa couplings of the neutral  scalars with respect to those 
of $H_{\rm SM}$ for this class of models~\cite{hunter}:
\begin{align}
g_{ht\bar{t}}  \ & = \  \cos\alpha/\sin\beta\; , \qquad 
g_{hb\bar{b}}\ =\ -\sin\alpha/\cos\beta\; , \qquad \nonumber\\
g_{Ht\bar{t}} \ & = \  \sin\alpha/\sin\beta\; , \qquad\! 
g_{Hb\bar{b}}\ =\ \cos\alpha/\cos\beta\; , \qquad \nonumber\\
g_{at\bar{t}} \ & = \  \cot\beta\; , \quad \qquad\qquad\!\!
g_{ab\bar{b}}\ =\ \tan\beta\; . 
\label{couplings}
\end{align}
Finally, we also write down the coupling of the charged scalar to the third-generation 
quarks~\cite{hunter}:
\begin{eqnarray}
g_{h^-t\bar{b}} \ = \ \frac{g}{2\sqrt 2
  M_W}[m_t\cot\beta(1+\gamma_5)+m_b\tan\beta(1-\gamma_5)]\; . 
\label{coup-hp}
\end{eqnarray}

\section{Two-loop RGEs in a General 2HDM}\label{app:RGE}

In  this  section,  we  present  the  two-loop  renormalization  group
equations  (RGEs) for  the general  2HDM given  by  (\ref{pot}). These
results were obtained using the general prescription given in~\cite{Machacek:1983tz}, 
as implemented in the public {\tt Mathematica} package {\tt  SARAH 4}~\cite{sarah}.

The two-loop RGEs for the $SU(3)_c$, $SU(2)_L$ and $U(1)_Y$ gauge couplings are respectively 
given by
\begin{align}
{\mathcal D}g_3  \ = \ & -\frac{7
  g_3^3}{16\pi^2}\ +\ \frac{g_3^3}{256\pi^4}
\left(-26 g_3^2 + \frac{9}{2}g_2^2 + \frac{11}{6}g'^2-2y_b^2-2y_t^2
  \right)\; ,\label{g3} \\  
{\mathcal D}g_2  \ = \ & -\frac{3 g_2^3}{16\pi^2
}\ +\ \frac{g_2^3}{256\pi^4}
\left( 12g_3^2+8g_2^2+2g'^2-\frac{3}{2}y_b^2-\frac{3}{2}y_t^2-\frac{1}{2}y_\tau^2
  \right)\;,\label{g2}\\ 
{\mathcal D}g' \ = \ &  \frac{7 g'^3}{16\pi^2}\  +\
\frac{g'^3}{256\pi^4}
\left( \frac{44}{3}g_3^2+6g_2^2+\frac{104}{9}g'^2-\frac{5}{6}y_b^2
- \frac{17}{6}y_t^2  - \frac{5}{2}y_\tau^2 \right)\; , \label{g1} 
\end{align}
where ${\mathcal D}\equiv d/d\ln \mu$ and $\mu$ is the usual  't-Hooft mass employed in the regularization
of  ultraviolet  divergences  in  loop  integrals. Note  that  at  the
one-loop level,  the gauge coupling RGEs  do not depend  on the Yukawa
and scalar couplings, whilst at two loops, they do depend on the quark
and lepton Yukawa couplings, for  which we have only kept the dominant
third-generation contributions.

Similarly for the Yukawa RGEs, we will only consider the third-generation Yukawa
couplings, such that for Type-II 2HDM, we have
\begin{eqnarray}
m_t \ = \ \frac{v_2}{\sqrt{2}} h_{2,33}^u \ \equiv
\ \frac{v_2}{\sqrt{2}} y_t \; , \qquad 
m_{b(\tau)} \ = \ \frac{v_{1}}{\sqrt{2}} h_{1,33}^{d(e)} \  \equiv
\ \frac{v_{1}}{\sqrt{2}} y_{b(\tau)}  \; . 
\end{eqnarray}
With  this approximation,  the third-generation Yukawa
coupling RGEs are given by
\begin{align}
{\mathcal D}y_t \  = \ & \frac{y_t}{16\pi^2 }\left(-8g_3^2-\frac{9}{4}g_2^2-\frac{17}{12}g'^2 + \frac{9}{2} y_t^2+\frac{1}{2}y_b^2 \right) \nonumber\\ 
& + \frac{y_t}{256\pi^4}\left[-108g_3^4-\frac{21}{4}g_2^4+\frac{1267}{216}g'^4+9g_2^2g_3^2-\frac{3}{4}g_2^2g'^2+\frac{19}{9}g_3^2g'^2 +6\lambda_2^2 +\lambda_3^2
 \right. \nonumber \displaybreak \\ 
& \left.+\lambda_3\lambda_4 +\lambda_4^2 +\frac{3}{2}(\lambda_5^2+\lambda_6^2+3\lambda_7^2) 
+\left(\frac{16}{3}g_3^2+\frac{33}{16}g_2^2 -\frac{41}{144}g'^2-2\lambda_3+2\lambda_4 \right)y_b^2 
\right. \nonumber\\
& \left.+\left(36g_3^2+\frac{225}{16}g_2^2+\frac{131}{16}g'^2-12\lambda_2\right)y_t^2 -\frac{5}{2}y_b^4-12y_t^4-\frac{5}{2}y_b^2y_t^2-\frac{3}{4}y_b^2y_\tau^2
\right],   \\
{\mathcal D} y_b \  = \ & \frac{y_b}{16\pi^2 }\left(-8g_3^2-\frac{9}{4}g_2^2-\frac{5}{12}g'^2 + \frac{9}{2}y_b^2+\frac{1}{2}y_t^2+y_\tau^2\right) \nonumber\\ 
& +\frac{y_b}{256\pi^4}\left[-108g_3^4-\frac{21}{4}g_2^4-\frac{113}{216}g'^4+9g_2^2g_3^2-\frac{9}{4}g_2^2g'^2+\frac{31}{9}g_3^2g'^2+6\lambda_1^2+\lambda_3^2\right.\nonumber\\
&\left. +\lambda_3\lambda_4+\lambda_4^2+\frac{3}{2}(\lambda_5^2+3\lambda_6^2+\lambda_7^2)+\left(\frac{16}{3}g_3^2+\frac{33}{16}g_2^2-\frac{53}{144}g'^2-2\lambda_3+2\lambda_4\right)y_t^2
\right.\nonumber\\
&\left. +\left(\frac{15}{8}g_2^2+\frac{25}{8}g'^2\right)y_\tau^2 
+\left(36g_3^2+\frac{225}{16}g_2^2+\frac{79}{16}g'^2-12\lambda_1 \right)y_b^2
\right.\nonumber\\
& \left. 
-12y_b^4-\frac{9}{4}y_\tau^4-\frac{5}{2}y_t^4-\frac{5}{2}y_b^2y_t^2-\frac{9}{4}y_b^2y_\tau^2
\right]\; , \\ 
{\mathcal D}y_\tau \  = \ & \frac{y_\tau}{16\pi^2}\left(-\frac{9}{4}g_2^2-\frac{15}{4}g'^2+3y_b^2+\frac{5}{2}y_\tau^2 \right)\nonumber \\ 
& +\frac{y_\tau}{256\pi^4}\left[-\frac{21}{4}g_2^4+\frac{161}{8}g'^4+\frac{9}{4}g_2^2g'^2  +6\lambda_1^2+\lambda_3^2+\lambda_3\lambda_4+\lambda_4^2\right. \nonumber \\ 
& \left. +\frac{3}{2}(\lambda_5^2+3\lambda_6^2+\lambda_7^2) +\left(20g_3^2+\frac{45}{8}g_2^2+\frac{25}{24}g'^2\right)y_b^2
\right.\nonumber\\
&\left. 
+\left(\frac{165}{16}g_2^2+\frac{179}{16}g'^2-12\lambda_1\right)y_\tau^2 
-\frac{27}{4}y_b^4-3y_\tau^4-\frac{27}{4}y_b^2y_\tau^2-\frac{9}{4}y_b^2y_t^2
\right]\; .
\end{align}
Similarly, the two-loop RGEs for the VEVs are given by 
\begin{align}
{\mathcal D}v_1 \ = \ &
\frac{v_1}{16\pi^2}\left[\frac{3}{4}(3g_2^2+g'^2)
- 3y_b^2-y_\tau^2\right]\nonumber\\ 
& +\frac{v_1}{256\pi^4}\left[\frac{435}{32}g_2^4-\frac{149}{32}g'^4-\frac{3}{16}g_2^2g'^2-6\lambda_1^2-\lambda_3^2-\lambda_3\lambda_4-\lambda_4^2
\right.\nonumber\\
&\left. -\frac{3}{2}(\lambda_5^2+3\lambda_6^2+\lambda_7^2) -\left(20g_3^2+\frac{45}{8}g_2^2+\frac{25}{24}g'^2\right)y_b^2-\left(\frac{15}{8}g_2^2+\frac{25}{8}g'^2\right)y_\tau^2 \right.\nonumber\\
& \left. +\frac{27}{4}y_b^4+\frac{9}{4}y_\tau^4+\frac{9}{4}y_b^2y_t^2\right]
-\frac{3 v_2}{512 \pi^4}
\bigg[(2\lambda_1+\lambda_{345})\lambda_6
+ (2\lambda_2+\lambda_{345})\lambda_7\bigg]\; , \\ 
{\mathcal D}v_2 \ = \ &
\frac{v_2}{16\pi^2}\left[\frac{3}{4}(3g_2^2+g'^2)-3y_t^2\right]\nonumber\\ 
& +\frac{v_2}{256\pi^4}\left[\frac{435}{32}g_2^4-\frac{149}{32}g'^4-\frac{3}{16}g_2^2g'^2-6\lambda_2^2-\lambda_3^2-\lambda_3\lambda_4-\lambda_4^2
\right. \nonumber\\
& \left. -\frac{3}{2}(\lambda^2_5+\lambda^2_6+3\lambda^2_7) 
- \left(20g_3^2+\frac{45}{8}g_2^2+\frac{85}{24}g'^2\right)y_t^2+\frac{9}{4}y_b^2y_t^2+\frac{27}{4}y_t^4
\right]\nonumber\\
& -\frac{3 v_1}{512 \pi^4}
\bigg[(2\lambda_1+\lambda_{345})\lambda_6+(2\lambda_2+\lambda_{345})\lambda_7
\bigg]\; .
\end{align}

The  two-loop RGEs  for all the scalar  quartic couplings appearing in (\ref{pot}) in the
Type-II 2HDM are given by
\begin{align}
{\mathcal D}\lambda_1 \ = \ & \frac{1}{16\pi^2 }\Bigg[\frac{3}{8}(3g_2^4+g'^4+2g_2^2g'^2) - 3\lambda_1(3g_2^2+g'^2)+24\lambda_1^2 + 2\lambda_3^2 + 2\lambda_3\lambda_4 + \lambda_4^2 
\nonumber\\
& \qquad \qquad 
+ \lambda_5^2 + 12\lambda_6^2 
+ 4\lambda_1 (3y_b^2+y_\tau^2)  - 6 y_b^4 - 2y_\tau^4\Bigg] \nonumber \\ 
& 
+ \frac{1}{256\pi^4}\Bigg[\frac{1}{16}\left(291 g_2^6-101 g_2^4g'^2-191g_2^2g'^4-131g'^6\right)
\nonumber\\
& -\frac{1}{8}(51g_2^4-78g_2^2g'^2-217g'^4)\lambda_1
+\frac{5}{2}(3g_2^4+g'^4)\lambda_3+\frac{5}{4}(3g_2^4+2g_2^2g'^2+g'^4)\lambda_4
\nonumber\\
& 
+(3 g_2^2+g'^2)(36\lambda_1^2+4\lambda_3^2+4\lambda_3\lambda_4
+\lambda_4^2+18\lambda_6^2)
+g'^2(\lambda_4^2-\lambda_5^2) -312\lambda_1^3 
\nonumber\\
& 
-8\lambda_3^3-6\lambda_4^3-20\lambda_1\lambda_3\lambda_4-4(5\lambda_1+3\lambda_4)
\lambda_3^2
-4(3\lambda_1+4\lambda_3)\lambda_4^2
\nonumber\\
& 
-2(7\lambda_1+10\lambda_3+11\lambda_4)\lambda_5^2 -2(159\lambda_1+33\lambda_3+35\lambda_4+37\lambda_5)\lambda_6^2 
\nonumber\\
& 
-4(9\lambda_3+7\lambda_4+5\lambda_5)\lambda_6\lambda_7
+2(3\lambda_1-9\lambda_3-7\lambda_4-5\lambda_5)\lambda_7^2
\nonumber \\ 
& 
-\left\{\frac{9}{4}g_2^4-\frac{9}{2}g_2^2g'^2-\frac{5}{4}g'^4
-\left(\frac{45}{2}g_2^2+80g_3^2+\frac{25}{6}g'^2\right)\lambda_1 
+36(4\lambda_1^2+\lambda_6^2) \right\}y_b^2 
\nonumber\\
& 
-\Big(32g_3^2-\frac{4}{3}g'^2+3\lambda_1\Big)y_b^4
 -6(2\lambda_3^2+2\lambda_3\lambda_4+\lambda_4^2
+\lambda_5^2+6\lambda_6^2)y_t^2
\nonumber\\
& 
-\left\{\frac{3}{4}g_2^4-\frac{11}{2}g_2^2g'^2
+\frac{25}{4}g'^4-\frac{5}{2}(3g_2^2+5g'^2)\lambda_1
+12(4\lambda_1^2
+\lambda_6^2)\right\}y_\tau^2
\nonumber\\
& 
-(4g'^2+\lambda_1)y_\tau^4-9\lambda_1y_b^2y_t^2+6y_t^2y_b^4
+30y_b^6+10y_\tau^6 
\Bigg]\; , 
 \\
{\mathcal D}\lambda_2 \ =\ & \frac{1}{16\pi^2 }\Bigg[\frac{3}{8}(3g_2^4+g'^4+2g_2^2g'^2) - 3\lambda_2(3g_2^2+g'^2)+24\lambda_2^2 + 2\lambda_3^2 + 2\lambda_3\lambda_4 
\nonumber\\
& \qquad \qquad  + \lambda_4^2 + \lambda_5^2 + 12\lambda_7^2 
 + 12\lambda_2 y_t^2 - 6 y_t^4\Bigg]\nonumber\\
& +\frac{1}{256\pi^4}\Bigg[\frac{1}{16}\left(291g_2^6-101g_2^4g'^2-191g_2^2g'^4-131g'^6\right)
\nonumber\\
& 
-\frac{1}{8}\left(51g_2^4-78g_2^2g'^2-217g'^4  \right)\lambda_2
+\frac{5}{2}(3g_2^4+g'^4)\lambda_3 +\frac{5}{4}(3g_2^4+2g_2^2g'^2+g'^4)\lambda_4
\nonumber\\
& 
+(3g_2^2+g'^2)(36\lambda_2^2+4\lambda_3^2+4\lambda_3\lambda_4+\lambda_4^2+18\lambda_7^2)
+g'^2(\lambda_4^2-\lambda_5^2)
-312\lambda_2^3 
\nonumber \\ 
& 
-8\lambda_3^3 -6\lambda_4^3-20\lambda_2\lambda_3\lambda_4-4(5\lambda_2+3\lambda_4)\lambda_3^2
-4(3\lambda_2+4\lambda_3)\lambda_4^2
\nonumber\\
& 
-2(7\lambda_2+10\lambda_3+11\lambda_4)\lambda_5^2 +2(3\lambda_2-9\lambda_3-7\lambda_4-5\lambda_5)\lambda_6^2
\nonumber\\
& 
-4(9\lambda_3+7\lambda_4+5\lambda_5)\lambda_6\lambda_7
-2(159\lambda_2+33\lambda_3+35\lambda_4+37\lambda_5)\lambda_7^2 
\nonumber\\
& 
-6(2\lambda_3^2+2\lambda_3\lambda_4+\lambda_4^2+\lambda_5^2+6\lambda_7^2)y_b^2
-\left\{\frac{9}{4}g_2^4-\frac{21}{2}g_2^2g'^2+\frac{19}{4}g'^4 \right.
\nonumber \\
& \left.
-\left(\frac{45}{2}g_2^2+80g_3^2+\frac{85}{6}g'^2\right)\lambda_2+36(4\lambda_2^2
+\lambda_7^2)\right\}y_t^2-9\lambda_2y_b^2y_t^2+6y_b^2y_t^4
\nonumber\\
&
-\left(32g_3^2+\frac{8}{3}g'^2+3\lambda_2\right)y_t^4+30y_t^6-2(2\lambda_3^2+2\lambda_3\lambda_4+\lambda_4^2+\lambda_5^2+
6\lambda_7^2)y_\tau^2\Bigg]\; ,  \displaybreak  \\
{\mathcal D}\lambda_3 \ = \ & \frac{1}{16\pi^2 }\Bigg[\frac{3}{4}(3g_2^4+g'^4-2g_2^2g'^2)
-3\lambda_3(3g_2^2+g'^2)+4(\lambda_1+\lambda_2)(3\lambda_3+\lambda_4) + 4\lambda_3^2
\nonumber \\
& \quad + 2(\lambda_4^2 + \lambda_5^2) + 4(\lambda_6^2+\lambda_7^2+4\lambda_6\lambda_7)+2\lambda_3(3y_b^2+y_\tau^2+3y_t^2)-12y_b^2y_t^2
\Bigg] 
\nonumber  \\ 
&  +\frac{1}{256\pi^4}\Bigg[\frac{1}{8}\left(291g_2^6+11g_2^4g'^2+101g_2^2g'^4-131g'^6\right)
\nonumber\\
& 
+\frac{5}{2}\left(9g_2^4-2g_2^2g'^2+3g'^4 \right)(\lambda_1+\lambda_2)
-\frac{1}{8}(111g_2^4-22g_2^2g'^2-197g'^4)\lambda_3
\nonumber\\ &
+2(3g_2^2+g'^2)[12(\lambda_1+\lambda_2)\lambda_3+\lambda_3^2+\lambda_4^2]-4(\lambda_1^2+\lambda_2^2)(15\lambda_3+4\lambda_4)
\nonumber\\
& 
 -4(\lambda_1+\lambda_2)(18\lambda_3^2+7\lambda_4^2+8\lambda_3\lambda_4+9\lambda_5^2)
-12(\lambda_3^3+\lambda_4^3+g_2^2\lambda_3\lambda_4)  
\nonumber\\
&
+\left(\frac{15}{2}g_2^4-3g_2^2g'^2+\frac{5}{2}g'^4\right)\lambda_4
+4(9g_2^2+2g'^2)(\lambda_1+\lambda_2)\lambda_4
\nonumber \\ 
&
-4\lambda_3\lambda_4(\lambda_3+4\lambda_4)
-2(9\lambda_3+22\lambda_4)\lambda_5^2
-4g'^2(\lambda_4^2-\lambda_5^2)+2g'^2(\lambda_6^2+\lambda_7^2)
\nonumber\\ 
& 
-4(31\lambda_1+11\lambda_2)\lambda_6^2-4(15\lambda_3 +17\lambda_4+17\lambda_5)(\lambda_6^2 +\lambda_7^2)+4(27g_2^2+8g'^2) \lambda_6\lambda_7 
\nonumber\\
&  -8[11(\lambda_1+\lambda_2+2\lambda_3+\lambda_4)+9\lambda_5] \lambda_6\lambda_7 
 - 4 (11\lambda_1+31\lambda_2)\lambda_7^2  
\nonumber\\
&  -\left\{ \frac{1}{4}(9g_2^4-5g'^4+18g_2^2g'^2)-\big(40g_3^2+\frac{45}{4}g_2^2+\frac{25}{12}g'^2\big)\lambda_3\right\}y_b^2
\nonumber\\
& -\frac{1}{4}\Big\{3g_2^4+25g'^4+22g_2^2g'^2-5(3g_2^2+5g'^2)\lambda_3\Big\}y_\tau^2
\nonumber\\
&  -2\left\{2 \lambda_3^2 + \lambda_4^2 + 
  4 \lambda_1 (3 \lambda_3 + \lambda_4) + \lambda_5^2 + 
  4 \lambda_6^2 + 8 \lambda_6\lambda_7\right\}(3y_b^2+y_\tau^2)
\nonumber\\
& 
 -\Big\{\frac{9}{4}g_2^4+\frac{21}{2}g_2^2g'^2+\frac{19}{4}g'^4-\Big(40g_3^2+\frac{45}{4}g_2^2+\frac{85}{12}g'^2\Big)\lambda_3 
\nonumber\\
& 
+6(12\lambda_2\lambda_3+2\lambda_3^2+4\lambda_2\lambda_4+\lambda_4^2+\lambda_5^2+8\lambda_6\lambda_7+4\lambda_7^2)\Big\}y_t^2
\nonumber\\
& 
-\big(64g_3^2+\frac{4}{3}g'^2-15\lambda_3\big)y_b^2y_t^2+36y_b^4y_t^2
-\frac{9}{2}\lambda_3[3(y_b^4+y_t^4)+y_\tau^4]+36y_b^2y_t^4 
\Bigg]\; ,  \\
{\mathcal D}\lambda_4 \ = \ & \frac{1}{16\pi^2 }\Bigg[3g_2^2g'^2 -3\lambda_4(3g_2^2+g'^2) 
+4(\lambda_1+\lambda_2+2\lambda_3)\lambda_4 + 4\lambda_4^2 + 8\lambda_5^2 
\nonumber \\
& \qquad \quad + 10(\lambda_6^2+\lambda_7^2) + 4\lambda_6\lambda_7 +2\lambda_4\big\{3(y_b^2+y_t^2)+y_\tau^2\big\}+12y_b^2y_t^2 \Bigg] \nonumber \\ 
& +\frac{1}{256\pi^4} \Bigg[-14g_2^4g'^2-\frac{73}{2}g_2^2g'^4+2g_2^2g'^2\big\{5(\lambda_1+\lambda_2)+\lambda_3\big\}
 \nonumber\\
&
-\frac{1}{8}\big(231g_2^4-102g_2^2g'^2-157g'^4\big)\lambda_4 +4\big\{2g'^2(\lambda_1+\lambda_2)
-7(\lambda_1^2+\lambda_2^2+\lambda_3^2)\big\}\lambda_4
 \nonumber\\
& 
+4(9g_2^2+g'^2)\lambda_3\lambda_4-40(\lambda_1+\lambda_2)(2\lambda_3\lambda_4+\lambda_4^2) 
+2(9g_2^2+4g'^2)\lambda_4^2-28\lambda_3\lambda_4^2
 \nonumber \\ 
& +2(27g_2^2+8g'^2)\lambda_5^2-48(\lambda_1+\lambda_2+\lambda_3)\lambda_5^2-26\lambda_4\lambda_5^2
+8g'^2\lambda_6\lambda_7
 \nonumber\\
& +2\big\{27g_2^2+7g'^2-2(18\lambda_3+17\lambda_4+20\lambda_5)\big\}(\lambda_6^2+\lambda_7^2)
-4(37\lambda_1+5\lambda_2)\lambda_6^2
\nonumber\\ &
-8\{5(\lambda_1+\lambda_2+2\lambda_3+4\lambda_4)+12\lambda_5\}\lambda_6\lambda_7 
 -4(5\lambda_1+37\lambda_2)\lambda_7^2 
\nonumber  \\ 
&+\Big\{9g_2^2g'^2+\big(40g_3^2 +\frac{45}{4}g_2^2+\frac{25}{12}g'^2 \big)\lambda_4
-12[2(\lambda_1+\lambda_3)\lambda_4+\lambda_4^2+2\lambda_5^2
\nonumber \displaybreak
\\ &  
+5\lambda_6^2+\lambda_6\lambda_7]\Big\}y_b^2  
+\Big\{21g_2^2g'^2
+\big(40g_3^2+\frac{45}{4}g_2^2+\frac{85}{12}g'^2\big)\lambda_4
-12[2(\lambda_2+\lambda_3)\lambda_4
\nonumber\\
& 
+\lambda_4^2
+2\lambda_5^2+\lambda_6\lambda_7+5\lambda_7^2]\Big\}y_t^2
+\Big(64g_3^2+\frac{4}{3}g'^2-24\lambda_3-33\lambda_4  \Big)y_b^2y_t^2
\nonumber \\
&
 +\Big\{11g_2^2g'^2
+\frac{5}{4}\left(3g_2^2+5g'^2  \right)\lambda_4
-4[2(\lambda_1+\lambda_3)\lambda_4+\lambda_4^2+2\lambda_5^2
\nonumber\\
&
+5\lambda_6^2+\lambda_6\lambda_7]\Big\}y_\tau^2
-\frac{9}{2}\lambda_4\big\{3(y_b^4+y_t^4)+ y_\tau^4\big\}-24y_b^2y_t^2(y_b^2+y_t^2)
\Bigg]\; , \\ 
{\mathcal D}\lambda_5 \ = \ & \frac{1}{16\pi^2 }\Bigg[- 3\lambda_5(3g_2^2+g'^2) + 4(\lambda_1+\lambda_2+2\lambda_3+3\lambda_4)\lambda_5 
  + 10(\lambda_6^2+\lambda_7^2)
\nonumber \\ &
\qquad \qquad  + 4\lambda_6\lambda_7 
+ 2\lambda_5 \big\{3(y_b^2+y_t^2)+y_\tau^2\big\}
\Bigg] \nonumber \\ 
& +\frac{1}{256\pi^4}\Bigg[-\frac{1}{8}\Big(231g_2^4-38g_2^2g'^2-157g'^4\Big)\lambda_5 
+4(9g_2^2+4g'^2)\lambda_3\lambda_5
\nonumber\\ & 
-4\Big\{g'^2(\lambda_1+\lambda_2)+7(\lambda_1^2+\lambda_2^2+\lambda_3^2)\Big\}\lambda_5
-8(\lambda_1+\lambda_2)(10\lambda_3+11\lambda_4)\lambda_5
 \nonumber \\
&
+4\{6(3g_2^2+g'^2)-19\lambda_3-8\lambda_4\}\lambda_4\lambda_5
+6\lambda_5^3-4(37\lambda_1+5\lambda_2)\lambda_6^2
\nonumber\\
&
-4(5\lambda_1+37\lambda_2)\lambda_7^2
+2\{27g_2^2+10g'^2-2(18\lambda_3+19\lambda_4+18\lambda_5)\}(\lambda_6^2+\lambda_7^2)
\nonumber\\
&
-4\{g'^2+10(\lambda_1+\lambda_2+2\lambda_3)
+22\lambda_4+42\lambda_5\}\lambda_6\lambda_7
\nonumber \\ &
+\Big\{\big(40g_3^2+\frac{45}{4}g_2^2+\frac{25}{12}g'^2\big)\lambda_5
-12[2(\lambda_1+\lambda_3)\lambda_5+3\lambda_4\lambda_5+5\lambda_6^2+\lambda_6\lambda_7]\Big\} y_b^2 
\nonumber\\ &
+\Big\{\big(40g_3^2+\frac{45}{4}g_2^2+\frac{85}{12}g'^2\big)\lambda_5-12[2(\lambda_2+\lambda_3)\lambda_5+3\lambda_4\lambda_5+\lambda_6\lambda_7+5\lambda_7^2\Big\}y_t^2
\nonumber\\ &
+\Big\{\big(\frac{15}{4}g_2^2+\frac{25}{4}g'^2\big)\lambda_5-8(\lambda_1+\lambda_3)\lambda_5-12\lambda_4\lambda_5-20\lambda_6^2-4\lambda_6\lambda_7\Big\}y_\tau^2
\nonumber \\ &
-\frac{1}{2}\lambda_5\{3(y_b^4+y_t^4)+y_\tau^4\}-33\lambda_5y_b^2y_t^2
\Bigg]\; ,  \\ 
{\mathcal D}\lambda_6 \ = \ &  \frac{1}{16\pi^2 }\Bigg[-3\lambda_6(3g_2^2+g'^2)+2(12\lambda_1+3\lambda_3+4\lambda_4)\lambda_6+2(3\lambda_3+2\lambda_4)\lambda_7 
\nonumber \\ &
\qquad \qquad + 10\lambda_5\lambda_6+2\lambda_5\lambda_7
 +3\lambda_6(3y_b^2+y_t^2+y_\tau^2)\Bigg] \nonumber \\ 
& +\frac{1}{256\pi^4}\Bigg[   
-\frac{1}{8}(141g_2^4-58g_2^2g'^2-187g'^4)\lambda_6+6(3g_2^2+g'^2)(6\lambda_1+\lambda_3)\lambda_6
 \nonumber\\
&
-6(53\lambda_1^2-\lambda_2^2)\lambda_6
-4(33\lambda_1+9\lambda_2+8\lambda_3)\lambda_3\lambda_6
+2(18g_2^2+5g'^2)\lambda_4\lambda_6
 \nonumber\\
&
-2(70\lambda_1+14\lambda_2+34\lambda_3+17\lambda_4)\lambda_4\lambda_6
+2(27g_2^2+10g'^2)\lambda_5\lambda_6
 \nonumber\\
&
-4(37\lambda_1+5\lambda_2+18\lambda_3+19\lambda_4+9\lambda_5)\lambda_5\lambda_6
-111\lambda_6^3-42\lambda_7^3
\nonumber \\ &
+\frac{5}{4}(9g_2^4+2g_2^2g'^2+3g'^4)\lambda_7
+12(3g_2^2+g'^2)\lambda_3\lambda_7-36(\lambda_1+\lambda_2+\lambda_3)\lambda_3\lambda_7
\nonumber\\
& +2(9g_2^2+4g'^2)\lambda_4\lambda_7-2\{14(\lambda_1+\lambda_2+2\lambda_3)
+17\lambda_4\}\lambda_4\lambda_7
\nonumber  \\
& -2\{g'^2+10(\lambda_1+\lambda_2+2\lambda_3)+22\lambda_4+21\lambda_5\}\lambda_5\lambda_7
-3(42\lambda_6+11\lambda_7)\lambda_6\lambda_7
 \nonumber \\ 
&
+\Big\{60g_3^2+\frac{135}{8}g_2^2+\frac{25}{8}g'^2 
-6(24\lambda_1+3\lambda_3+4\lambda_4+5\lambda_5)
 \Big\}\lambda_6y_b^2 
 \nonumber 
 \displaybreak \\ 
&
+
\Big\{20g_3^2+\frac{45}{8}g_2^2+\frac{85}{24}g'^2-6(3\lambda_3+4\lambda_4+5\lambda_5)\Big\}\lambda_6y_t^2
 \nonumber\\
& 
+\Big\{\frac{15}{8}(3g_2^2+5g'^2)-2(24\lambda_1+3\lambda_3+4\lambda_4+5\lambda_5)\Big\}\lambda_6 y_\tau^2
\nonumber\\ &
-12(3\lambda_3+2\lambda_4+\lambda_5)\lambda_7y_t^2-\frac{1}{4}(27y_t^4+33y_b^4+11y_\tau^4)\lambda_6-21\lambda_6y_b^2y_t^2
\Bigg]\; , \\
{\mathcal D}\lambda_7 \ = \ & \frac{1}{16\pi^2 }\Bigg[-3\lambda_7(3g_2^2+g'^2)+2(12\lambda_2+3\lambda_3+4\lambda_4)\lambda_7+2(3\lambda_3+2\lambda_4)\lambda_6 
 \nonumber\\
&
\qquad \qquad + 10\lambda_5\lambda_7+2\lambda_5\lambda_6+\lambda_7(3y_b^2+9y_t^2+y_\tau^2)\Bigg]\nonumber  \\ 
& +\frac{1}{256\pi^4}\Bigg[ 
\frac{5}{4}(9g_2^4+2g_2^2g'^2+3g'^4)\lambda_6+12(3g_2^2+g'^2)\lambda_3\lambda_6
\nonumber\\
& 
-36(\lambda_1+\lambda_2+\lambda_3)\lambda_3\lambda_6
+2(9g_2^2+4g'^2)\lambda_4\lambda_6-28(\lambda_1+\lambda_2+2\lambda_3)\lambda_4\lambda_6
\nonumber \\ &
-34\lambda_4^2\lambda_6-2g'^2\lambda_5\lambda_6-4\{5(\lambda_1+\lambda_2+2\lambda_3)+11\lambda_4\}\lambda_5\lambda_6-42(\lambda_5^2+\lambda_6^2)\lambda_6
\nonumber\\
& -\frac{1}{8}(141g_2^4-58g_2^2g'^2-187g'^4)\lambda_7+6\lambda_1^2\lambda_7+36(3g_2^2
+g'^2)\lambda_2\lambda_7-318\lambda^2_2\lambda_7
\nonumber\\
&
+6(3g_2^2+g'^2)\lambda_3\lambda_7-12(3\lambda_1+11\lambda_2)\lambda_3\lambda_7
-32\lambda^2_3\lambda_7
+2(18g_2^2+5g'^2)\lambda_4\lambda_7
 \nonumber\\
&-4(7\lambda_1+35\lambda_2+17\lambda_3)\lambda_4\lambda_7
-34\lambda^2_4\lambda_7
+2(27g_2^2+10g'^2)\lambda_5\lambda_7
\nonumber\\
&
-4(5\lambda_1+37\lambda_2+18\lambda_3+19\lambda_4)\lambda_5\lambda_7-36\lambda^2_5\lambda_7-33\lambda_6^2\lambda_7
\nonumber \\ &
-126\lambda_6\lambda_7^2-111\lambda_7^3-12(3\lambda_3 +2\lambda_4+\lambda_5)\lambda_6y_b^2
\nonumber \\ &
+\Big\{20g_3^2+\frac{45}{8}g_2^2+\frac{25}{24}g'^2-6(3\lambda_3+4\lambda_4+5\lambda_5)\Big\}\lambda_7y_b^2
 \nonumber\\
&
+ \Big\{60g_3^2+ \frac{135}{8}g_2^2+\frac{85}{8}g'^2-6(24\lambda_2+3\lambda_3+4\lambda_4+5\lambda_5)
 \Big\}\lambda_7 y_t^2
 \nonumber  \\
&
-4(3\lambda_3+2\lambda_4+\lambda_5)\lambda_6y_\tau^2  
+\Big\{\frac{5}{8}(3g_2^2+5g'^2)-2(3\lambda_3+4\lambda_4+5\lambda_5)\Big\}\lambda_7 y_\tau^2
\nonumber\\ &
-\frac{1}{4}(33y_t^4+27y_b^4+9 y_\tau^4)\lambda_7-21\lambda_7y_b^2y_t^2 
\Bigg]\; . 
\end{align}
The two-loop RGE for the soft mass parameter is given by 
\begin{align}
{\mathcal D}(m^2_{12}) \ = \ & \frac{1}{16\pi^2}\Bigg[ 
-\frac{3}{2}(3g_2^2+g'^2)m^2_{12}+2(\lambda_3+2\lambda_4+3\lambda_5)m^2_{12}
\nonumber \\ 
& \qquad \qquad +2(3y_b^2+3y_t^2+y_\tau^2)m^2_{12}+12(\lambda_6\mu^2_1+\lambda_7\mu^2_2) \Bigg]
\nonumber \\ 
& +\frac{1}{256\pi^4}\Bigg[
-\frac{1}{16}(243g_2^4-30g_2^2g'^2-153g'^4)m^2_{12}+3\big\{2\lambda_1^2+\lambda_2^2)+\lambda_5^2
\nonumber \\ &
+4(\lambda_6^2+\lambda_7^2)\big\}m^2_{12}
+4(3g_2^2+g'^2)(\lambda_3+2\lambda_4+3\lambda_5) m^2_{12}
\nonumber 
\\ &
-12(\lambda_1+\lambda_2)\lambda_{345} 
m^2_{12}-6(\lambda_3\lambda_4+2\lambda_3\lambda_5+2\lambda_4\lambda_5+6\lambda_6\lambda_7)m^2_{12}
\nonumber  
\\ &
+\Big\{20g_3^2+\frac{45}{8}g_2^2+\frac{25}{24}g'^2
-6(\lambda_3+2\lambda_4+3\lambda_5)\Big\}y_b^2m^2_{12}
\nonumber \displaybreak \\ &
+\Big\{20g_3^2+\frac{45}{8}g_2^2+\frac{85}{24}g'^2
-6(\lambda_3+2\lambda_4+3\lambda_5)\Big\}y_t^2m^2_{12}
\nonumber   \\ &
+\Big\{\frac{5}{8}(3g_2^2+5g'^2)
-2(\lambda_3+2\lambda_4+3\lambda_5)\Big\}y_\tau^2 m^2_{12}
\nonumber \\ &
+24(3g_2^2+g'^2)(\lambda_6\mu_1^2+\lambda_7\mu_2^2)
-72(\lambda_1\lambda_6\mu_1^2+\lambda_2\lambda_7\mu_2^2)
\nonumber \\ &
-12\lambda_{345}\{(2\lambda_6+\lambda_7)\mu_1^2+(\lambda_6+2\lambda_7)\mu_2^2\}
\nonumber \\ &
-24\{(3y_b^2+y_\tau^2)\lambda_6\mu^2_1+3y_t^2\lambda_7\mu_2^2\}
-\frac{9}{4}(3y_b^4+3y_t^4+y_\tau^4)m^2_{12}
\Bigg]
\end{align}
Finally, the two-loop RGE for the soft mass parameter is given by 
\begin{align}
{\mathcal D}(m^2_{12}) \ = \ & \frac{1}{16\pi^2}\Bigg[ 
-\frac{3}{2}(3g_2^2+g'^2)m^2_{12}+2(\lambda_3+2\lambda_4+3\lambda_5)m^2_{12}
\nonumber \\ 
& \qquad \qquad +2(3y_b^2+3y_t^2+y_\tau^2)m^2_{12}+12(\lambda_6\mu^2_1+\lambda_7\mu^2_2) \Bigg]
\nonumber \\ 
& +\frac{1}{256\pi^4}\Bigg[
-\frac{1}{16}(243g_2^4-30g_2^2g'^2-153g'^4)m^2_{12}+3\big\{2\lambda_1^2+\lambda_2^2)+\lambda_5^2
\nonumber \\ &
+4(\lambda_6^2+\lambda_7^2)\big\}m^2_{12}
+4(3g_2^2+g'^2)(\lambda_3+2\lambda_4+3\lambda_5) m^2_{12}
\nonumber 
\\ &
-12(\lambda_1+\lambda_2)\lambda_{345} 
m^2_{12}-6(\lambda_3\lambda_4+2\lambda_3\lambda_5+2\lambda_4\lambda_5+6\lambda_6\lambda_7)m^2_{12}
\nonumber \\ &
+\Big\{20g_3^2+\frac{45}{8}g_2^2+\frac{25}{24}g'^2
-6(\lambda_3+2\lambda_4+3\lambda_5)\Big\}y_b^2m^2_{12}
\nonumber \\ &
+\Big\{20g_3^2+\frac{45}{8}g_2^2+\frac{85}{24}g'^2
-6(\lambda_3+2\lambda_4+3\lambda_5)\Big\}y_t^2m^2_{12}
\nonumber \\ &
+\Big\{\frac{5}{8}(3g_2^2+5g'^2)
-2(\lambda_3+2\lambda_4+3\lambda_5)\Big\}y_\tau^2 m^2_{12}
\nonumber \\ &
+24(3g_2^2+g'^2)(\lambda_6\mu_1^2+\lambda_7\mu_2^2)
-72(\lambda_1\lambda_6\mu_1^2+\lambda_2\lambda_7\mu_2^2)
\nonumber \\ &
-12\lambda_{345}\{(2\lambda_6+\lambda_7)\mu_1^2+(\lambda_6+2\lambda_7)\mu_2^2\}
\nonumber \\ &
-24\{(3y_b^2+y_\tau^2)\lambda_6\mu^2_1+3y_t^2\lambda_7\mu_2^2\}
-\frac{9}{4}(3y_b^4+3y_t^4+y_\tau^4)m^2_{12}
\Bigg]
\end{align}
Note that the mass parameters $\mu_{1,2}^2$ are removed by the tadpole conditions:
\begin{eqnarray}
\frac{\partial V}{\partial v_1} = 0 &=& -\mu_1^2 v_1-m_{12}^2 v_2 +\frac{1}{2}\left[2\lambda_1 v_1^3 + \lambda_{345}v_1v_2^2 + 3\lambda_6v_1^2v_2+\lambda_7v_2^3\right],\\
\frac{\partial V}{\partial v_2} = 0 &=& -\mu_2^2 v_2-m_{12}^2 v_1 +\frac{1}{2}\left[2\lambda_2 v_2^3 + \lambda_{345}v_2v_1^2 + \lambda_6 v_1^3+3\lambda_7v_2v_1^2\right],
\end{eqnarray}  
and hence, it is not necessary to write down their RGEs explicitly. 

\end{document}